\documentclass[twocolumn, aps, amsmath, amssymb, prb, footinbib, citeautoscript, longbibliography, superscriptaddress, 10pt]{revtex4-2}

\usepackage[hmargin=1.57cm, vmargin=1.56cm]{geometry}
\usepackage{graphicx}					
\usepackage[]{hyperref}					
\hypersetup{colorlinks, allcolors=blue}		
\usepackage[lofdepth, lotdepth, caption=false]{subfig}		
\usepackage{comment}

\begin{document}
\title{Geometrical effects on the downstream conductance in\\quantum-Hall--superconductor hybrid systems}

\author{A.~David}
\author{J.~S.~Meyer}
\author{M.~Houzet}
\affiliation{Univ.~Grenoble Alpes, CEA, Grenoble INP, IRIG, Pheliqs, 38000 Grenoble, France}

\date{\today}

\begin{abstract}
We consider a quantum Hall (QH) region in contact with a superconductor (SC), i.e., a QH-SC junction. Due to successive Andreev reflections, the QH-SC interface hosts hybridized electron and hole edge states called chiral Andreev edge states (CAES). We theoretically study the transport properties of these CAES by using a microscopic, tight-binding model. We find that the transport properties strongly depend on the contact geometry and the value of the filling factor. We notice that it is necessary to add local barriers at the corners of the junction in order to reproduce such properties, when using effective one-dimensional models.
\end{abstract}

\maketitle

\section{Introduction\label{sec: introduction}}
{Combining systems displaying a quantum Hall effect and superconductors is a difficult task, as the magnetic field needed to realize the quantum Hall effect tends to suppress superconductivity. If successful, it leads to interesting phenomena as the superconductor may induce correlations in the chiral edge states of the quantum Hall system. In particular, the formation of so-called chiral Andreev edge states (CAES) has been predicted. Semiclassically, these CAES result from skipping orbits of electrons and holes involving  Andreev reflections at the quantum Hall-superconductor (QH-SC)  interface~\cite{takagaki1998, asano2000, chtchelkatchev2001, chtchelkatchev2007}. Quantummechanically, the edge states along that interface are described as hybridized electron and hole states~\cite{hoppe2000, zulicke2001, giazotto2005, khaymovich2010, ostaay2011}. Their use for topologically protected quantum computing was also considered~\cite{nayak2008, mong2014, clarke2014}.}

{A number of recent experiments have succeeded in creating QH-SC hybrid systems using either graphene~\cite{lee2017, zhao2020, gul2022, zhao2022} or InAS two-dimensional electron gas (2DEG)~\cite{hatefipour2022},  and observing evidence for CAES in the so-called downstream conductance. Namely, the downstream conductance measures the conversion of electrons into holes, involving the transfer of Cooper pairs into the superconductor along the interface. The larger the conversion probability, the smaller the downstream conductance and, in particular, it becomes negative when the conversion probability exceeds one half. While the experiments~\cite{lee2017, zhao2020, gul2022, zhao2022, hatefipour2022} did indeed measure negative downstream conductances, questions remain about the magnitude and the parameter dependence of the effect that do not match simple models: the observed signal is much smaller than expected. Furthermore, it shows either an irregular pattern~\cite{lee2017, zhao2020, gul2022} or remains roughly constant~\cite{hatefipour2022} when sweeping the field or the gate voltage, while simple models predict a regular oscillation. This stimulated further theoretical research. A suppression of the measured signal may be explained by the absorption of quasiparticles in the superconductor, for example, by subgap states in nearby vortices~\cite{zhao2020, manesco2021, kurilovich2022, schiller2022}, whereas the oscillations may be strongly affected by disorder~\cite{manesco2021, kurilovich2022}.}

{Here we explore a different aspect that has not been addressed before: the role of the geometry. Namely, the downstream conductance does not probe only the properties of the QH-SC interface, but also the scattering properties at the point where this interface meets the QH-vacuum interface. We find that these scattering probabilities strongly depend on the geometry of the contact region. In particular, a pronounced dependence of the angle between the QH-vacuum interface and the QH-SC interface is observed. Interestingly, this opens the possibility of creating asymmetric structures, where the angles are different on the two sides of the superconductor, that may display an enhanced overall electron-hole conversion probability. This may even lead to a situation where the downstream conductance becomes negative on average.}

{Note that to study the effect of geometry, a full two-dimensional description of the system is necessary -- simple one-dimensional models commonly used in the literature are not sufficient. Some aspects may be captured by using a generalized one-dimensional model, though there is no obvious way to determine its parameters.}

{The paper is organized as follows. In Sec.~\ref{sec: system and conductance formula}, we present the system and the downstream conductance formula based on edge state transport whose parameters have to be computed. To do so, we first use a two-dimensional model in Sec.~\ref{sec: 2D model}. In particular, we start by studying a continuous model in  Sec.~\ref{ssec:2Dc} that allows one to determine the properties of the edge states at an infinitely long interface. We then use a tight-binding model in Sec.~\ref{sec: tight-binding simulations} to obtain the scattering probabilities at the points where two different interfaces, i.e., QH-vacuum and QH-SC, meet. With these two ingredients, we have all that is needed to compute the downstream conductance. In Sec.~\ref{sec: 1D model}, we address the question whether the prior results may be obtained from an effective one-dimensional model. Further considerations on the role of additional nonchiral edge states and the effects of temperature can be found in Sec.~\ref{sec: limits of 1D model}, before we conclude in Sec.~\ref{sec: conclusion}. Some details were  relegated to the appendices.}


\section{System and conductance formula\label{sec: system and conductance formula}}

The conductance along the edge of  a system in the  quantum Hall regime can be attributed to the properties of its chiral edge states. We are interested in the regime, where one spin-degenerate Landau level is occupied in the quantum Hall region, i.e., there are two chiral edge states. Introducing particle-hole space to be able to incorporate superconductivity, we can describe one spin state as an electron state and the other spin state as a hole state.
While the chiral edge states along an edge with the vacuum are either pure electron or hole states, the CAES along an edge with a superconductor are a superposition of electron and hole components. In the following, we will call them quasielectron when their momentum at the Fermi level is negative and quasihole when their momentum at the Fermi level is positive.
As we will see below, for the system under consideration, this choice is in agreement with the pure electron and hole states obtained when Andreev processes are suppressed.

\begin{figure}\centering
\includegraphics[width=.9\linewidth]{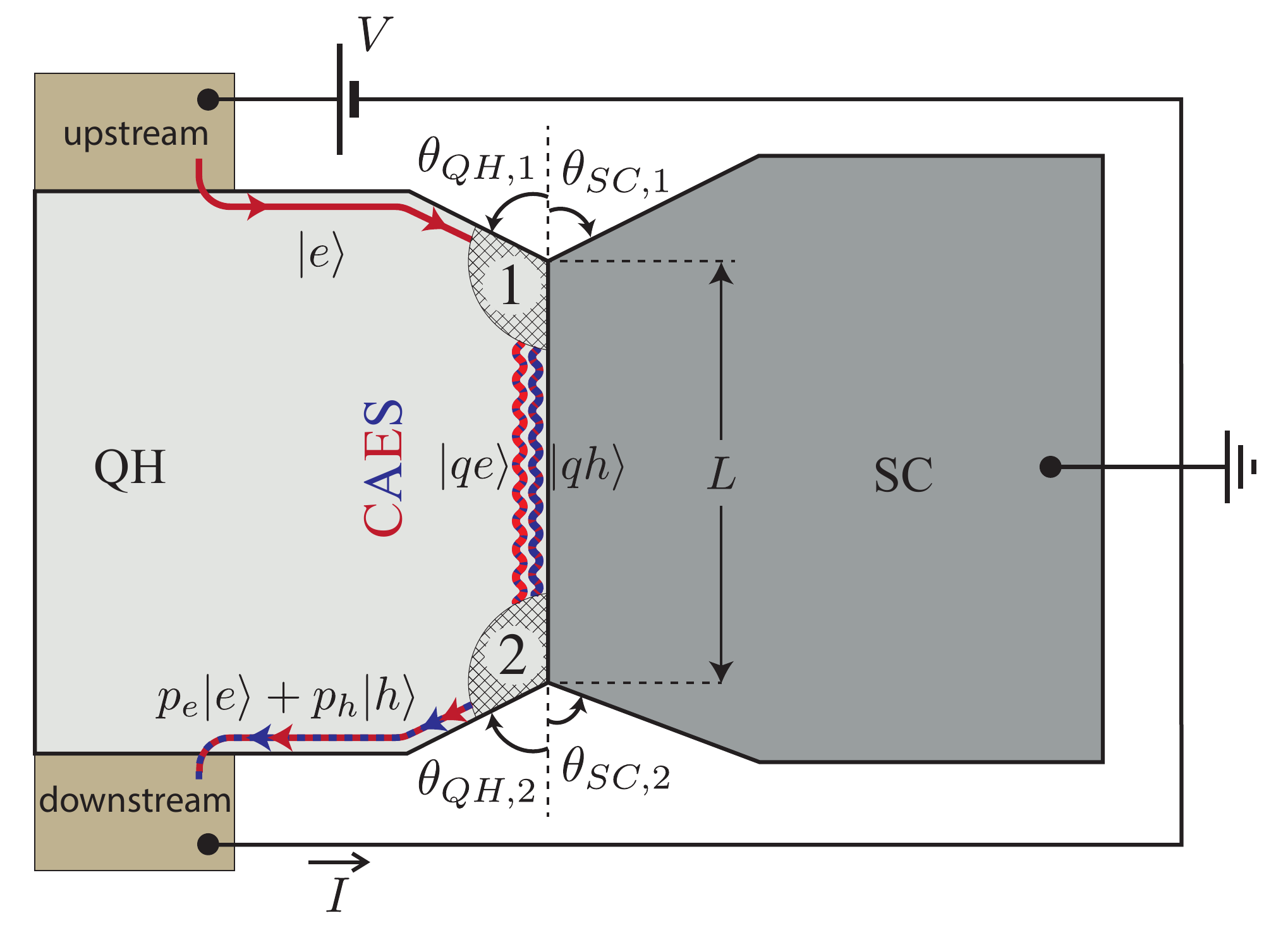}
\caption{QH-SC setup: the edge of the quantum Hall region is in contact with a grounded superconductor over a finite length $L$.  The geometry of the corners at the beginning and end of that region can be characterized by two angles each: $\theta_{QH,i}$ and $\theta_{SC,i}$.  Both the QH-vacuum and QH-SC interface host chiral edge states that can be probed by measuring the differential downstream conductance 
$G_d= \partial I/\partial V$, where $V$ is the voltage applied to the upstream reservoir and $I$ is the current flowing into the downstream reservoir. While (quasi)electron and (quasi)hole states have opposite directions of quasi-momentum along the interface, they have the same propagation direction.
A typical process contributing to $G_d$ is shown. An incoming electron $|e\rangle$ scatters at the first corner, propagates along the QH-SC interface as a superposition of quasielectron $|qe\rangle$ and quasihole $|qh\rangle$ CAES, then scatters at the second corner, and finally exits the superconductor in a superposition of electron $|e\rangle$ and hole $|h\rangle$. 
The hole probability $P_h=|p_h|^2$ of the outgoing state depends on the scattering processes at the corners as well as the interference of the CAES propagation along the QH-SC interface.}
\label{fig: qh_sc_junction}
\end{figure}

We want to study the situation where the edge of the quantum Hall system  is in contact with a superconductor over a region with finite length $L$ as shown in Fig.~\ref{fig: qh_sc_junction}. In that case, we can define a probability $P_h$ that an incoming electronlike state is transformed into an outgoing holelike state. It depends on the properties of the CAES along the QH-SC interface as well as the scattering amplitudes at the two corners, which begin and end that interface. Assuming ballistic propagation along the interface with a given material, the probability $P_h$ can be written as~\cite{khaymovich2010}
\begin{eqnarray}
P_h&=&\tau_1(1-\tau_2)+\tau_2(1-\tau_1)\nonumber\\
&&+2\sqrt{\tau_1 (1-\tau_2)\tau_2 (1-\tau_1)} \cos(2 k_0 L+\phi_{12}),
\label{eq-Ph1D}
\end{eqnarray}
where $\tau_1$ is the probability that the electron is converted into a quasihole at the beginning of the QH-SC interface, whereas $\tau_2$ is the probability that a quasielectron is converted into a hole at the end of the QH-SC interface. The second line describes the interference resulting from the fact that the particle may propagate along the QH-SC interface either as a quasielectron with momentum $-k_0$ or as a quasihole with momentum $+k_0$. The phase shift $\phi_{12}$ depends on the phases of the scattering amplitudes at the two corners. At zero temperature, the differential downstream conductance, $G_d(0)= \partial I/\partial V|_{V=0}$, where $V$ is the voltage applied to the upstream reservoir and $I$ is the current flowing into the downstream reservoir (see Fig.~\ref{fig: qh_sc_junction}), is directly related to the probability $P_h$ at the Fermi level, namely $G_d(0)=G_0(1-2P_h)$, where $G_0=2e^2/h$ is the conductance quantum. A negative downstream conductance is a clear signature of the Andreev conversion taking place at the QH-SC interface. Note that the average conductance is given as $\bar G_d=G_0\prod_{i=1,2}(1-2\tau_i)$. For $\tau_1=\tau_2$ it is limited to positive values, whereas $\tau_1\neq\tau_2$ allows one to realize $\bar G_d<0$. For completeness, let us mention that the maximal downstream conductance is $G_d^{\rm max}=G_0[1-2(\sqrt{\tau_1(1-\tau_2)}-\sqrt{\tau_2(1-\tau_1)}\,)^2]$, while the minimal 
downstream conductance is $G_d^{\rm min}=G_0[1-2(\sqrt{\tau_1(1-\tau_2)}+\sqrt{\tau_2(1-\tau_1)}\,)^2]$. In the symmetric case $\tau_1=\tau_2\equiv\tau$, this yields $G_d^{\rm max}=G_0$ and  $G_d^{\rm min}=G_0[1-8\tau(1-\tau)]$.

Thus, to model the experimentally measured  downstream conductance, we  need to determine $k_0$ as well as the probabilities $\tau_i$ associated with the contact points between the QH region, the vacuum, and the superconductor. In the following, we show that $k_0$ can be obtained semianalytically from a microscopic model of an infinite QH-SC interface. By contrast, there is no simple model for the probabilities $\tau_i$. We study their dependence on system parameters and, in particular, the geometry of the contact points using tight-binding simulations. To conclude, we compare with an effective 1D model.

\section{Two-dimensional model\label{sec: 2D model}}

\subsection{Continuum model of an infinite QH-SC interface\label{ssec:2Dc}}

We will consider an interface along the $y$ axis such that the region $x<0$ is in the quantum Hall regime whereas the region $x>0$ is a superconductor. The microscopic Hamiltonian can be written in the form
\begin{equation}
\label{eq: microscopic hamiltonian0}
H = 
\begin{pmatrix}
H_0 - \mu(x) 	&	\Delta(x) \\
\Delta^*(x)	& -H_0^*+ \mu(x)
\end{pmatrix}
\
\end{equation}
with ${\bf r}=(x,y)$ and
\begin{equation}
H_0=\frac{1}{2m} \left(-i{ \bf {\nabla}} - e {\bf A}(x)\right)^2 + V(x),
\label{eq: microscopic hamiltonian}
\end{equation}
using units where $\hbar=1$. Here, $\mu(x)=\mu_{QH}\Theta(-x)+\mu_{SC}\Theta(x)$ accounts for the drop of the chemical potential measured from the band bottom in the 2DEG and the superconductor, $\mu_{QH}$ and $\mu_{SC}$, respectively, $\Delta(x)=\Delta\Theta(x)$ is the superconducting order parameter with amplitude $\Delta$ (that we will choose to be real in the following), the potential $V(x) = V_0 \delta(x)$ with strength $V_0$ models an interface barrier, and $\Theta(x)$ is the Heaviside function. Note that we neglect self-consistency of the order parameter. Furthermore, we assume that the magnetic field in the superconductor is screened. Thus, choosing the Landau gauge that preserves translational invariance along the interface, we set ${\bf A}(x)=B x \Theta(-x)\hat{u}_y$. The wave functions can then be written in the form: 
\begin{equation}
\Psi({\bf r}) = \frac{e^{i k_y y}}{\sqrt{L_y}} \psi_{k_y}(x),
\end{equation}
where $L_y$ is the length of the system along the $y$-direction and $\psi_{k_y}$ is the transverse wave function associated with longitudinal wave vector $k_y$.
Following~\cite{blonder1982}, we can determine the CAES by writing the wave functions $\psi_{k_y}^{QH}(x)$ in the half-space $x<0$ and $\psi_{k_y}^{SC}(x)$ in the half-space $x>0$, and matching them at the interface to obtain an eigenstate of Eq.~\eqref{eq: microscopic hamiltonian0} at energy $E$.

In the QH region, one obtains
\begin{equation}
\psi_{k_y}^{QH}(x) = c^{QH}_+ \begin{pmatrix}1\\0\end{pmatrix}\chi_+(x) + c^{QH}_-\begin{pmatrix}0\\1\end{pmatrix}\chi_-(x)
\label{eq: qh wave function}
\end{equation}
with
\begin{align}
\chi_{\pm}(x) &= N_{\pm} U\left(-\frac{\mu_{QH} \pm E}{\omega_c}, -\frac{\sqrt{2}}{l_B}(x \mp k_y l_B^2)\right),
 \label{eq: qh parabolic cylinder functions} 
 \end{align}
where $U(a, z)$ are parabolic cylinder functions that vanish as $z\rightarrow-\infty$ (see Ref.~\cite{abramowitz1972} for the formal definition), and $N_\pm$ are normalization coefficients such that $\int_{-\infty}^0dx\;|\chi_\pm(x)|^2=1$. 
{Here we introduced the cyclotron frequency $\omega_c = eB/m$ and the magnetic length $l_B = 1/\sqrt{eB}$.}

Restricting ourselves to the regime  $|E| < \Delta$ ,  the wavefunctions in the SC region take the form~\cite{kulik1970}:
\begin{equation}
\psi_{k_y}^{SC}(x) = c^{SC}_{+}\frac1{\sqrt2} \begin{pmatrix}\gamma\\1\end{pmatrix}\phi(x) + c^{SC}_{-} \frac1{\sqrt2} \begin{pmatrix}\gamma^{*}\\1\end{pmatrix}\phi^{*}(x)
\label{eq: sc wave function}
\end{equation}
with $\phi(x) = \sqrt{2 \,\text{Im}\,q} \;e^{i q x}$,
\begin{align}
q^2 &= (k_{F}^{SC})^2 - k_y^2 + 2 i m\Delta\sqrt{1-\epsilon^2},
\end{align}
and $\gamma = \epsilon + i\sqrt{1-\epsilon^2}$, where $\epsilon=E/\Delta$ and $k_{F}^{SC} = \sqrt{2 m \mu_{SC}}$.

The matching procedure, $\psi_{k_y}^{QH}(0) = \psi_{k_y}^{SC}(0)$ and
$\partial_x\psi_{k_y}^{SC}(0) - \partial_x\psi_{k_y}^{QH}(0) = 2mV_0 \psi_{k_y}^{SC}(0) $,
yields the following secular equation for the energy $E(k_y)$~\cite{hoppe2000}:
\begin{align}
s(E, k_y) \equiv \,&G H \left(c^2 + d^2\right) + G' H' + d(G'H + GH') \nonumber \\
		&+ c \frac{\epsilon}{\sqrt{1 - \epsilon^2}}(G'H - GH')  = 0
\label{eq: secular equation}
\end{align}
with the shorthand notations
$c =\mathrm{Re}\,q$, $d = \mathrm{Im}\,q+ 2 m V_0$, 
$G = \chi_+(0)$, $G' = \chi_+'(0)$, $H = \chi_-(0)$, and $H' = \chi_-'(0)$, where the primes denote derivatives with respect to $x$.

\begin{figure}\centering
\includegraphics[width=0.7\linewidth]{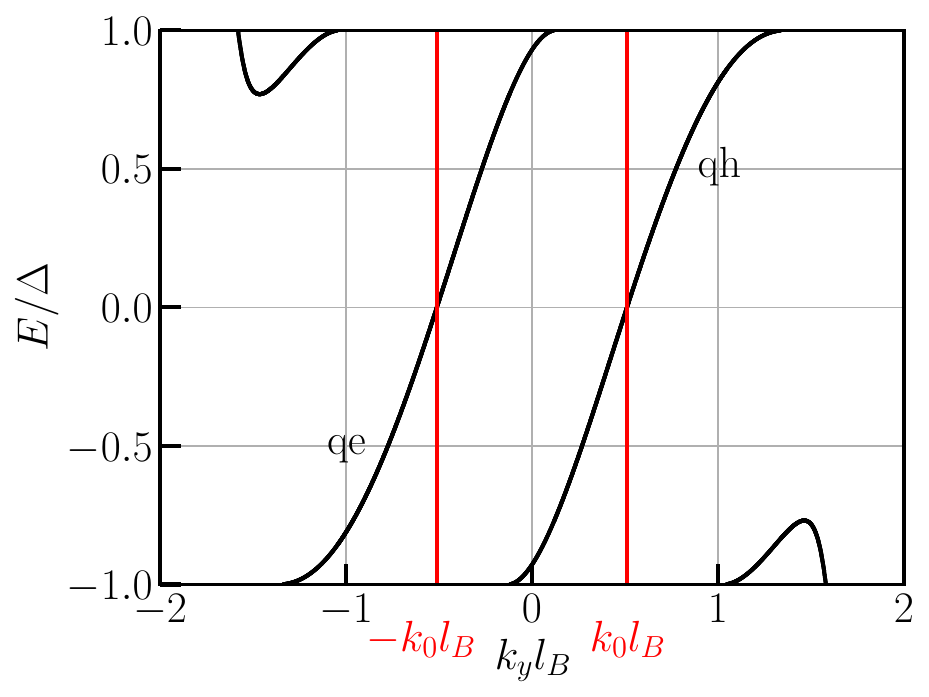}
\caption{Energy spectrum of the states along the QH-SC interface obtained from Eq.~\eqref{eq: secular equation}. The crossings of the CAES with the Fermi level are indicated by red lines. Here the parameters are  $\mu_{QH} = \mu_{SC} = 10\Delta$, $\nu=2.4$, and $V_0=0$. }
\label{fig: example of CAES spectrum}
\end{figure}

{When the filling factor $\nu\equiv  2\mu_{QH}/\omega_c$ is in the range $1<\nu<3$, the chemical potential lies between the first and second (spin-degenerate) Landau levels of the QH region, and one obtains a single pair of CAES. An example of the spectrum is shown in Fig.~\ref{fig: example of CAES spectrum}, 
where we considered an ideal interface, i.e., $\mu_{QH} = \mu_{SC}$ and $V_0 = 0$.   
At low energies, we see the two linearly dispersing CAES with energies  $E_\pm(k_y)=v_{\rm CAES}(k_y \pm k_0)$.}

\begin{figure}[!b]\centering
\includegraphics[width=0.7\linewidth]{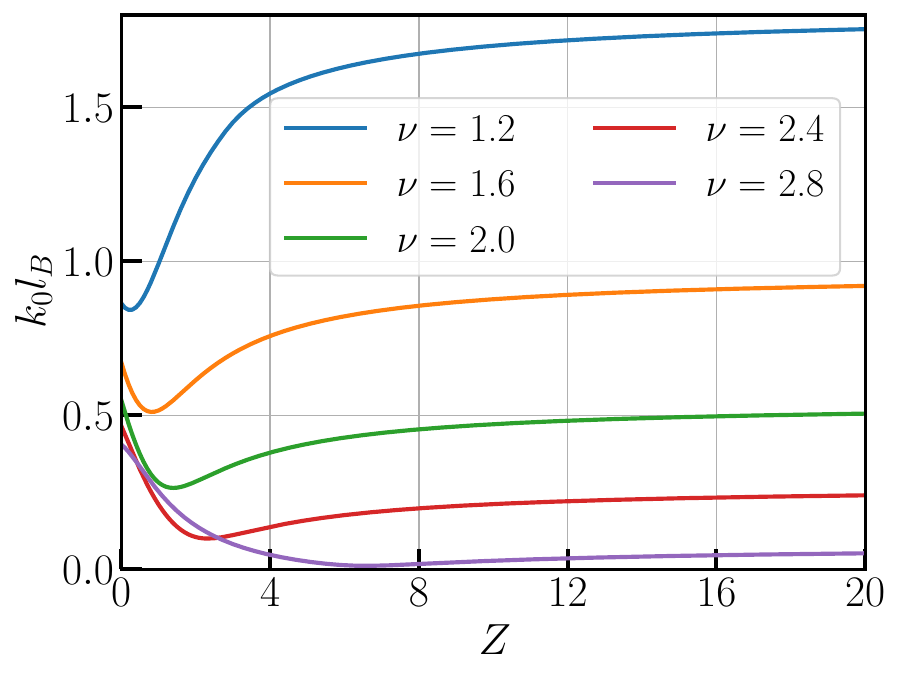}
\caption{Fermi momentum $k_0$ as a function of barrier strength $Z$ for various values of the filling factor $\nu$ at $\mu_{QH} = \mu_{SC} = 10\Delta$. Except for a small region of intermediate $Z$ and $\nu$ close to three, the momentum $k_0$ decreases with increasing $\nu$.}
\label{fig: k0 vs Z various fillings}
\end{figure}

The Fermi momentum $k_0$ appearing in the interference term for the downstream conductance, cf.~\eqref{eq-Ph1D}, can be obtained by solving $s(0,\mp k_0)=0$, which in general has to be done numerically. 
We observe that, as long as $\Delta \ll \mu_{QH},\mu_{SC}$, the momentum $k_0$ does not depend on $\Delta$. In the limit of a large interface barrier $Z\equiv 2mV_0\big/k_{F}^{QH}\gg1$ and $\nu\to3$, we find $k_0l_B\ll1$ and an analytical solution is possible, namely $k_0\approx(3-\nu)\sqrt\pi/4l_B$~\cite{ostaay2011}. In Fig.~\ref{fig: k0 vs Z various fillings}, we show the evolution of $k_0$ as a function of the barrier strength $Z$ for various values of the filling factor. Typically $k_0$ decreases with increasing $\nu$,  except for a small region of intermediate values of $Z$ and fillings $\nu$ close to three.

\begin{figure}[]\centering
\includegraphics[width=0.7\linewidth]{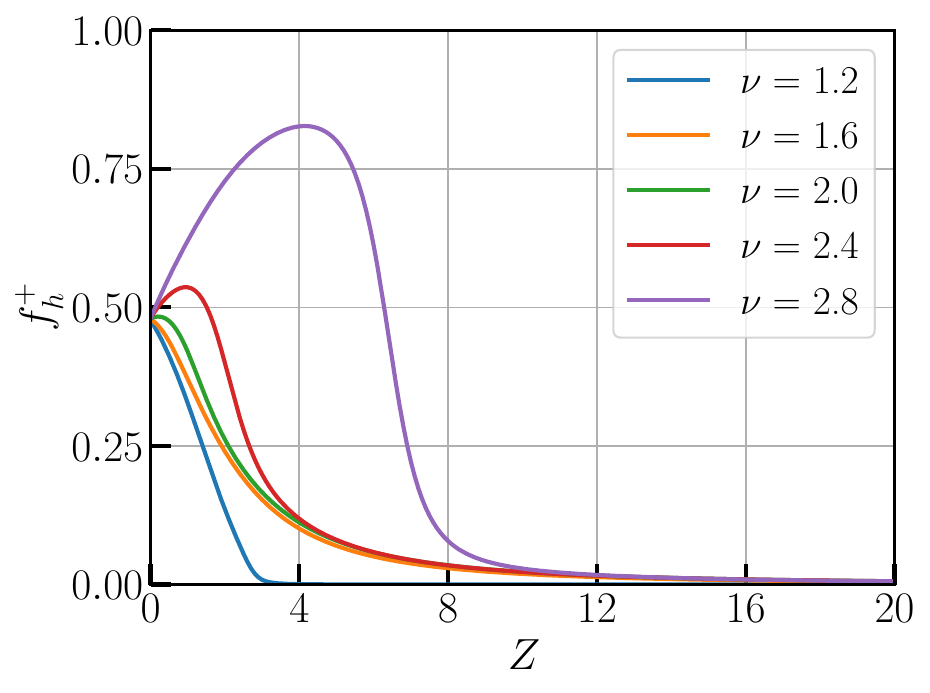}
\caption{Hole content $f_h^{+}$ of the quasielectron CAES versus the barrier's strength $Z$ for various values of the filling factor $\nu$ at $\mu_{QH} = \mu_{SC} = 10\Delta$. While at $Z=0$, the hole content is close to $1/2$:
it vanishes as $Z\gg1$. Interestingly, it is enhanced in an intermediate region for $\nu>2$. 
}
\label{fig: B vs Z various fillings}
\end{figure}

The velocity of the low-energy states is given as $v_{\rm CAES}= -\left.{\partial_{k_y} s(E, k_y)}/{\partial_E s(E, k_y)}\right|_{E=0, |k_y|=k_0}$. 
One may also compute the electron and hole content of the states, $\psi(x)=(\psi_e(x),\psi_h(x))$. We define
\begin{align}
f_h=\int_{-\infty}^\infty dx\;|\psi_h|^2=1-\int_{-\infty}^\infty dx\;|\psi_e|^2.
\end{align}
In particular, the result for the states at the Fermi level reads:
\begin{align}
\!\!\!f_h^+ =&1-\,c_0^2H_0^2\left(1\!+\!\frac{1}{4q_0"}\left(G_0^2+\frac{(g_0+q_0"G_0)^2}{|q_0|^2}\right)\right)\times\nonumber\\
&\,\times\left[g_0^2+c_0^2H_0^2\left(1\!+\!\frac1{2q_0"} \left(G_0^2+\frac{g_0^2}{c_0^2}\right)\right)\right]^{-1},
\end{align}
where $g_0=G_0'+d_0G_0$ and $q_0"={\rm Im}\,q_0$. Furthermore, the subscript 0 indicates that the previously introduced quantities have to be taken at $E=0$ and $k_y=-k_0$. {Particle-hole symmetry implies $f_h^-=1-f_h^+$. In the limit $Z\to\infty$, we recover pure electron and hole states, $f_h^+ = 0$ and $f_h^-=1$. By contrast, at $Z=0$ and in the limit $\Delta\to0$, one finds an equal repartition between electron and hole components, $f_h^+ = f_h^- = 1/2$. As an illustrative example, we represent the content $f_h^{+}$ of the quasielectron CAES as a function of the barrier strength $Z$ for various values of the filling factor $\nu$ in Fig. \ref{fig: B vs Z various fillings}.}

\subsection{{Tight-binding simulation and scattering probabilities}\label{sec: tight-binding simulations}}

We now turn to the scattering probabilities at the corners where the QH-vacuum interface and the QH-SC interface meet. In addition to the system parameters, this corner can be characterized by two angles as shown in Fig.~\ref{fig: qh_sc_junction}: the angle $\theta_{QH,i}$ that the QH-vacuum interface forms with the continuation of the QH-SC interface and the angle $\theta_{SC,i}$ that the SC-vacuum interface forms with the continuation of the QH-SC interface. 
{To ensure that there is no overlap, the angles must satisfy $\theta_{QH,i} + \theta_{SC,i} > 0$.}
To compute the scattering probabilities $\tau_i$ as a function of these angles and system parameters, we perform tight-binding simulations with a discretized version of the Hamiltonian \eqref{eq: microscopic hamiltonian} on a square lattice 
using the Kwant software~\cite{groth2014}. 
{
Introducing the Nambu spinor $\Psi_i = (c_i, c_i^{\dagger})^T$, where $c_i^{\dagger}$ ($c_i$) is the operator that creates (annihilates) an electron at the position $\mathbf{r}_{i}=(x_i,y_i)$, the second-quantized tight-binding Hamiltonian reads:
\begin{align}
\mathcal{H}_{TB} = &\sum_i \psi_i^\dagger \left[\left(4t - \mu_i + V_i\right)\sigma_z + \Delta_i\sigma_x \right]\psi_i \nonumber\\
		&+ \sum_{\langle i, j \rangle} \psi_i^\dagger \left[t e^{i\phi_{ij} \sigma_z}\sigma_z\right]\psi_j,
\label{eq: TB hamiltonian}
\end{align}
where $\sigma_{x/z}$ are Pauli matrices in Nambu space, and $\langle i, j \rangle$ denotes pairs of nearest neighbor sites. 
The barrier potential is given as $V_i=V_0 \delta_{x_i,0} \Theta\left(\frac L2 \!- \!|y_i|\right)$, where $\delta_{i,j}$ is the Kronecker delta. 
In the QH region, $\mu_i=\mu_{QH}$ and $\Delta_i=0$, whereas in the SC region, $\mu_i=\mu_{SC}$ and $\Delta_i=\Delta$. 
Using a Peierls substitution, the hopping matrix element $t = 1/(2ma^2)$, where $a$ is the lattice spacing, acquires a field-dependent phase~\cite{datta1995}
\begin{align}
\phi_{ij} &= -\frac{\pi B}{\phi_0} (x_i + x_j)(y_j - y_i) \theta\left(-\frac{x_i + x_j}{2}\right)
\end{align}
with $\phi_0$ the flux quantum. This lattice model matches the continuum model as long as the hopping energy is the largest energy scale, $\Delta,\mu_{QH}, \mu_{SC} \ll  t$. We further make realistic assumptions $\Delta \ll \mu_{QH} \leq \mu_{SC}$.}

\begin{figure}[!tb]\centering
\subfloat[\label{fig: tau vs theta_sc}]{\includegraphics[width=0.7\linewidth]{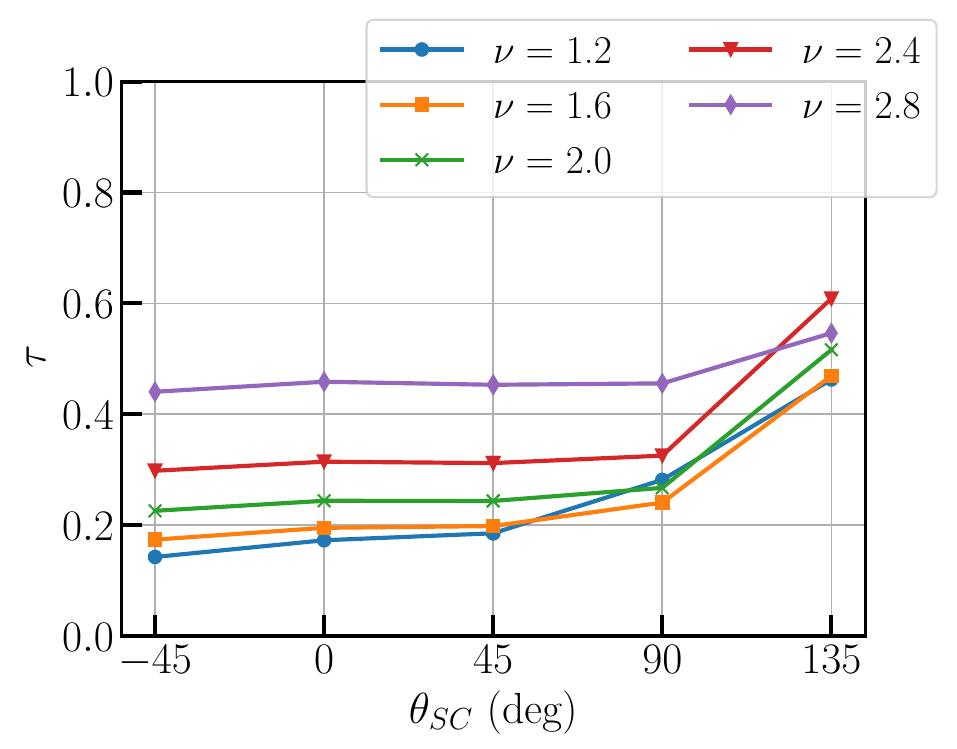}}

\subfloat[\label{fig: tau vs theta_qh}]{\includegraphics[width=0.7\linewidth]{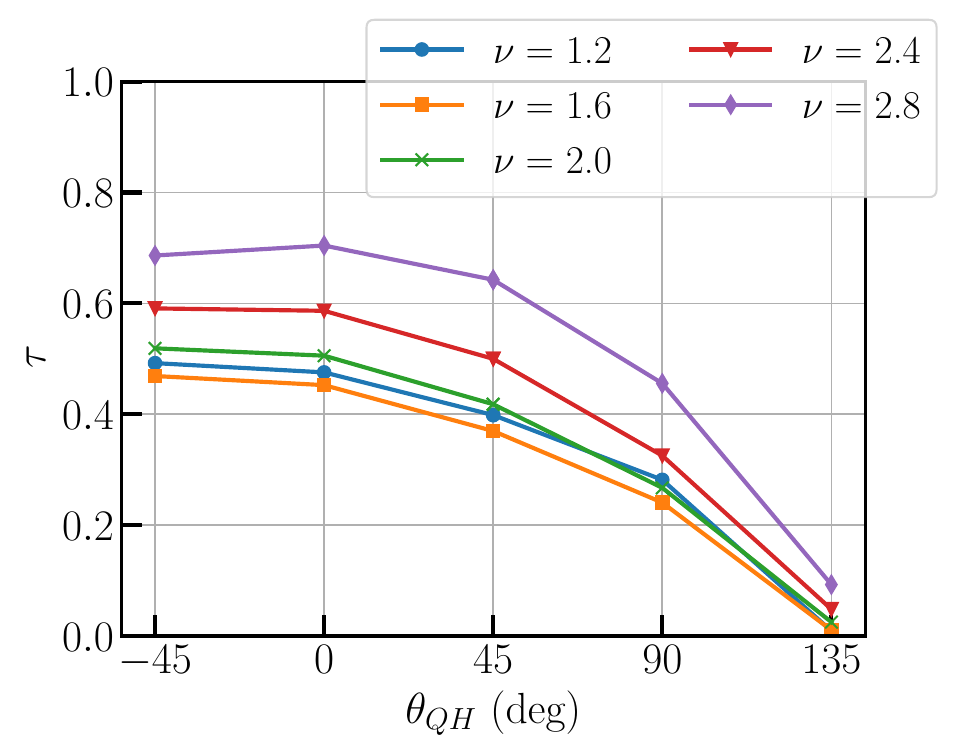}}
\caption{Conversion probability $\tau$ for various values of the filling factor $\nu$ as a function of (a) the SC angle $\theta_{SC}$ with $\theta_{QH} = 90^{\circ}$ and (b) the QH angle $\theta_{QH}$ with $\theta_{SC} = 90^{\circ}$. 
The parameters are $\mu_{QH} = \mu_{SC} = 10\Delta$ and $Z=0$. {To minimize lattice effects, we only show commensurate angles. The solid lines are a guide to the eye.}
}
\label{fig: tau vs theta various fillings}
\end{figure}

\begin{figure}[!b]\centering
\includegraphics[width=0.9\linewidth]{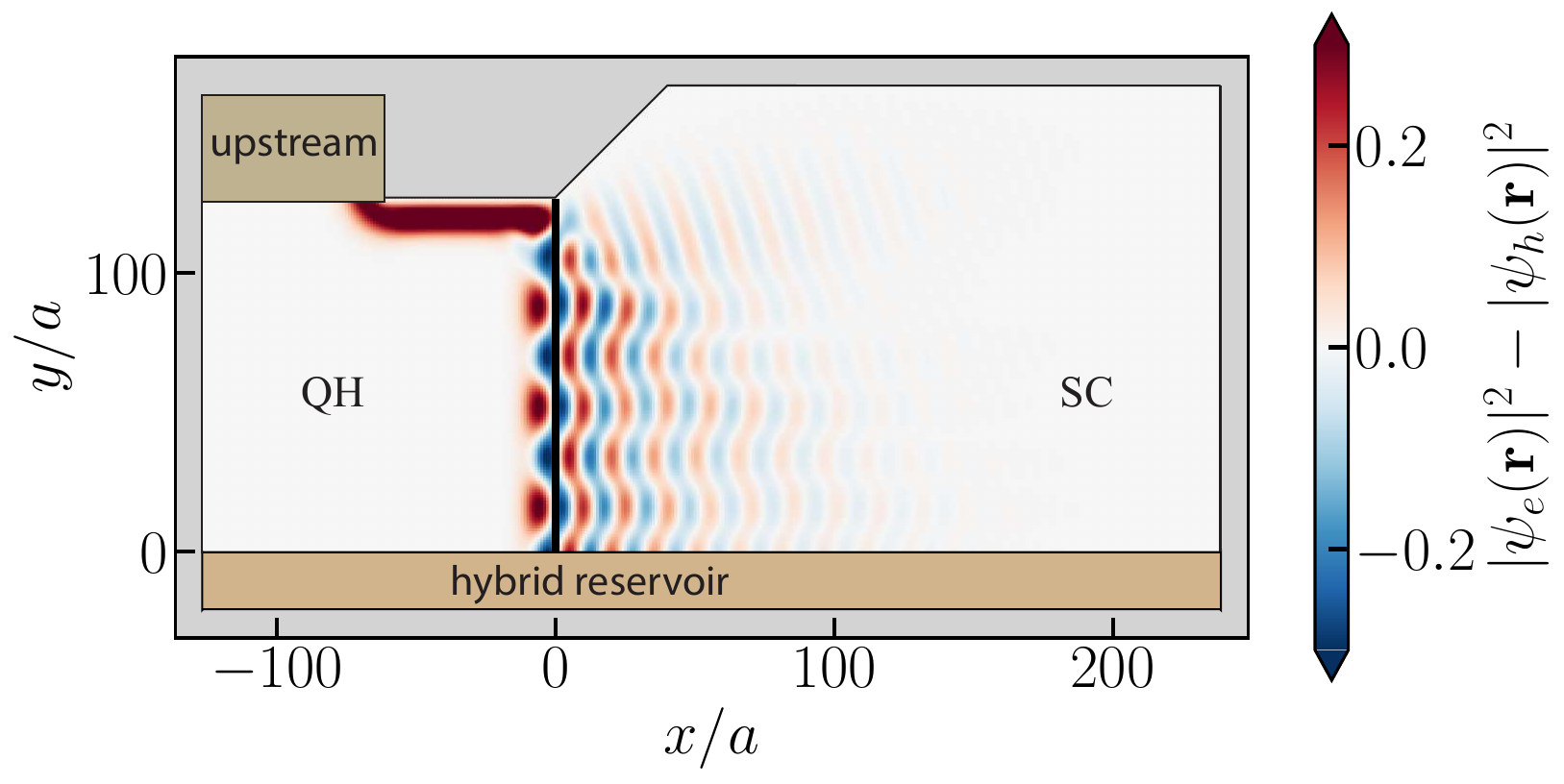}
\caption{Probability density $|\psi_e(\mathbf{r})|^2 - |\psi_h(\mathbf{r})|^2$ of an incoming electron state for $\theta_{SC}=45^{\circ}$ and $\theta_{QH} = 90^{\circ}$. 
The interference of CAES along the QH-SC interface (black line) can be clearly seen. Note that the wave function does not have any weight in the vicinity of the SC-vacuum boundary. 
The parameters are $\nu=2$, $\mu_{QH} = \mu_{SC} = 10\Delta$, and $Z=0$.}
\label{fig: density e-h}
\end{figure}

As the conversion probability from electron to quasihole at the first corner is equal to the conversion probability from quasi{electron to hole at the second corner when parameters are chosen the same~\cite{khaymovich2010}, $\tau_1(\theta_{QH},\theta_{SC})=\tau_2(\theta_{QH},\theta_{SC})\equiv\tau(\theta_{QH},\theta_{SC})$, 
it is sufficient to simulate the first QH-SC corner. The python code is available on Zenodo~\cite{david2023}. When not specified, we set $t = 1$ and $\mu_{SC} =  t/20$. 

Figure~\ref{fig: tau vs theta various fillings}  shows the dependence of $\tau$ on angles for $\mu_{QH} = \mu_{SC} = 10\Delta$. In Fig.~\ref{fig: tau vs theta_sc}, $\theta_{QH}=90^\circ$ is fixed while $\theta_{SC}$ varies. We see a weak dependence of $\tau$ on $\theta_{SC}$ for angles up to $90^\circ$.  This is not surprising as the propagation of the chiral edge states does not involve the SC-vacuum interface. The residual effect of $\theta_{SC}$ on the scattering probability is due to the modified decay of the edge state wave function into the bulk in the vicinity of the corner. As shown in Appendix~\ref{app: gap}, in this regime $\tau$ also shows a more pronounced dependence on the value of $\Delta$ that controls the decay length in the superconductor.
We illustrate this in Fig.~\ref{fig: density e-h} by plotting the probability density $|\psi_e(\mathbf{r})|^2 - |\psi_h(\mathbf{r})|^2$ of an incoming electron state; it can be seen that it is vanishingly small at angles $<90^\circ$ within the SC region.
By contrast, $\tau$ decreases as $\theta_{QH}$ is increased. This is shown in Fig.~\ref{fig: tau vs theta_qh}, where $\theta_{SC}=90^\circ$ is fixed while $\theta_{QH}$ varies.} 
The stronger sensitivity of $\tau$ on $\theta_{QH}$ can be understood as stemming from the fact that {this angle directly determines the propagation direction of the edge state and thus the projection of the momentum of the incoming state onto the direction of the interface.}

%

\begin{figure}[!tb]\centering
\includegraphics[width=0.7\linewidth]{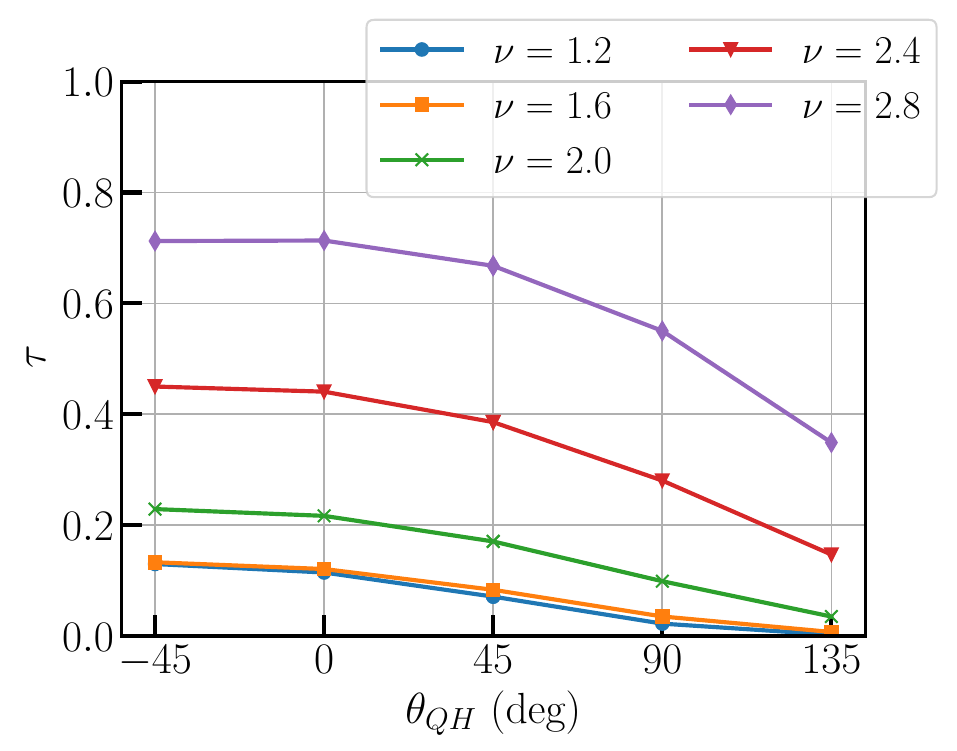}
\caption{Conversion probability $\tau$ versus the angle $\theta_{QH}$ for a nonideal interface at various values of $\nu$. Here $\theta_{SC} = 90^{\circ}$, $\mu_{SC} = 2\mu_{QH}=20\Delta$, and $Z=0.7$. As in Fig.~\ref{fig: tau vs theta various fillings}, we only show commensurate angles{, and the solid lines are a guide to the eye.}}
\label{fig: tau vs theta non-ideal interface}
\end{figure}

A more realistic interface is obtained when allowing for different values of $\mu_{QH}$ and $\mu_{SC}$, as well as for an interface barrier $Z\neq0$. 
As an example, in Fig.~\ref{fig: tau vs theta non-ideal interface} we show the evolution of $\tau$ as a function of $\theta_{QH}$ with $\mu_{SC} = 2\mu_{QH}$ and $Z = 0.7$. The behavior is qualitatively similar though the variation with the angle is less pronounced. The stronger variation with $\nu$ reflects the stronger variation of $f_h^+$ at intermediate values of $Z$, shown in Fig.~\ref{fig: B vs Z various fillings}. 

\section{One-dimensional model\label{sec: 1D model}}

Effective one-dimensional models are very useful to obtain a qualitative understanding of the edge state physics.
They have been extensively used in  recent works~\cite{zhao2020, hatefipour2022, kurilovich2022, schiller2022, gul2022} 
to describe the CAES. In this section, we address the question of how to incorporate the effects discussed in previous sections into such an effective model.

The starting point is the one-dimensional Bogoliubov-de Gennes Hamiltonian,
\begin{align}
\!\!\!\tilde{H} = \begin{pmatrix} -\frac i2\{{v}(y),\partial_y\}- {\mu}(y)\!\!\! & \tilde{\Delta}(y) \\ \tilde{\Delta}^{*}(y) & \!\!\!-\frac i2\{{v}(y),\partial_y\} +{\mu}(y) \end{pmatrix},
\label{eq: effective 1D Hamiltonian} 
\end{align}
where $y$ denotes the coordinate along the QH edge, $\tilde\Delta(y)$ are the induced superconducting correlations, $v(y)$ is the edge state velocity in the absence of superconducting correlations, and $\mu(y)$ is an effective chemical potential. Furthermore, $\{.,.\}$ is the anticommutator.

Choosing all the parameters to be independent of $y$ allows one to extract the zero-energy momentum $k_0$, the velocity $v_{\rm CAES}$, as well as the hole content $f_h^\pm$ of the CAES as introduced in Sec.~\ref{ssec:2Dc}. Diagonalizing $\tilde H$, one finds $E_\pm(k_y) = vk_y\pm\sqrt{\mu^2+\tilde\Delta^2}$ and $f_h^\pm=\left(1\pm\mu/\sqrt{\mu^2+\tilde\Delta^2}\right)/2$. To match the results of Sec.~\ref{ssec:2Dc}, we thus set $v=v_{\rm CAES}$,
\begin{eqnarray}
\mu&=&- v_{\rm CAES}k_0(1-2f_h^+),\label{eq-mu}\\
\tilde\Delta&=&v_{\rm CAES}k_0\sqrt{f_h^+(1-f_h^+)}.
\end{eqnarray}
The simplest model often used to describe scattering at the corner consists of choosing a step function for the induced correlations, $\tilde\Delta(y)=\tilde\Delta\Theta(y)$. Matching of the wave functions at the position of the step, $y=0$, directly yields
the conversion probability of an electron into a quasihole: 
\begin{equation}
\tau_0=f_h^+.
\end{equation}
This clearly is not sufficient to correctly describe the scattering---if only because it doesn't depend on the geometry of the contact point. Furthermore, it can be shown that choosing a different velocity $v_{\rm vac}\neq v$ and/or effective chemical potential $\tilde\mu_{\rm vac}\neq \tilde\mu$ for the QH-vacuum interface at $y<0$ does not modify this result. To obtain a conversion probability $\tau\neq\tau_0$, one needs to include a spatial variation of the induced correlations $\tilde\Delta(y)$ in the vicinity of $y=0$. 

We thus consider a more general model with a barrier region, $-L_b/2<y<L_b/2$, characterized by the parameters $v(y)=v_b$, $\mu(y)=\mu_b$ and $\tilde\Delta(y)=\Delta_be^{i\phi_b}$. Note that the relative  superconducting phase between the barrier and the bulk is allowed as time-reversal symmetry is broken by the applied field. 
Solving the Schrödinger equation in the three regions (QH-vacuum interface at $y<-L_b/2$, barrier, and QH-SC interface at $y>L_b/2$), matching the solutions at $y=\pm L_b/2$, and solving the resulting system, we obtain:
\begin{eqnarray}
\tau 
&=& \left(\sqrt{\tau_0}\cos\beta_b
+ \sqrt{1 - \tau_0}\sin\beta_b\right)^2\\
&&-4 \sqrt{\tau_0(1-\tau_0)} \sin\beta_b\cos\beta_b \cos^2\frac{\phi_b-\delta_b}2\nonumber
\label{eq-effective_tau}
\end{eqnarray}
with $\alpha_b=\sqrt{\mu_b^2 + \Delta_b^2}L_b/v_b$, $\sin\beta_b= \sin\alpha_b\,\Delta_b/\sqrt{\mu_b^2 + \Delta_b^2}$, and $\tan\delta_b=\cot\alpha_b\sqrt{\mu_b^2 + \Delta_b^2}/\mu_b$. {If $\Delta_b\neq0$, $\tau\neq\tau_0$ is possible, and the model has sufficient parameters to obtain an arbitrary value of $\tau$ for a given $\tau_0$.} Thus, in principal, this effective one-dimensional model can be used to describe an arbitrary geometry. However, there is no straightforward way to estimate parameters. As a consequence, a full two-dimensional model is necessary to determine the downstream conductance even in simple geometries.

\begin{figure}[!t]\centering
\includegraphics[width=0.7\linewidth]{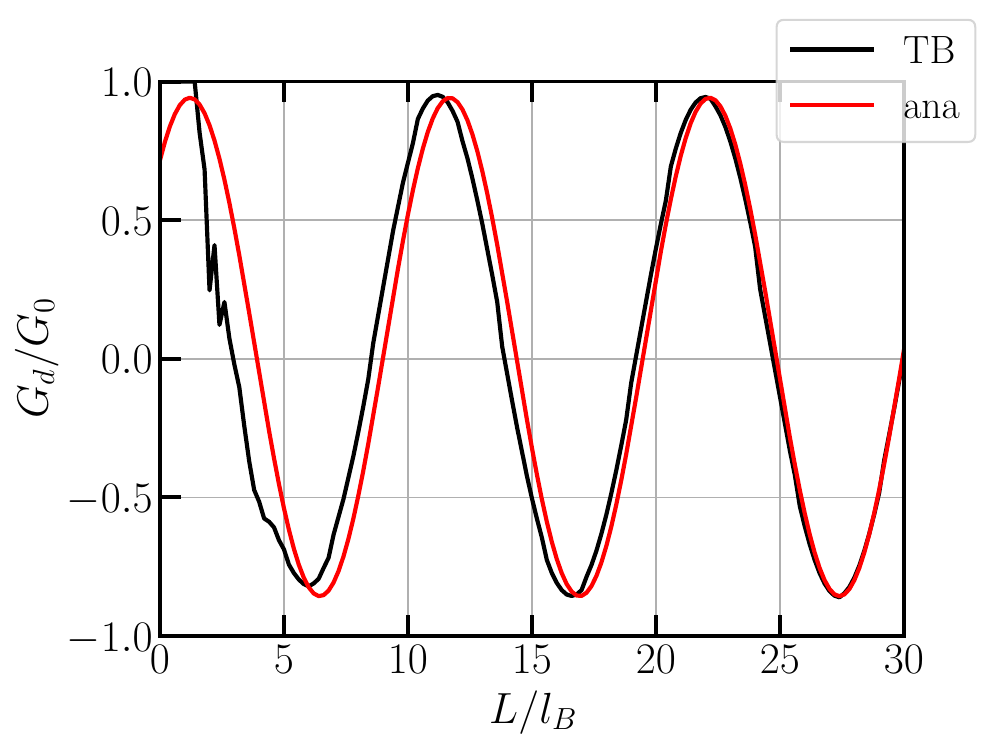}
\caption{Conductance oscillations as a function of length $L$ of an asymmetric junction, $\theta_{QH,1} = 0$ and $\theta_{QH,2} = 90^{\circ}$ whereas $\theta_{SC,1}=\theta_{SC,2}=90^\circ$. We compare a full tight-binding simulation (TB) with the results of an effective one-dimensional model, where the parameters have been chosen as discussed in Sec.~\ref{sec: 1D model}. Here, $\nu = 2.8$,  $\mu_{SC} = 2\mu_{QH}=20\Delta$ and $Z=0.7$. The scattering phase $\phi_{12}$ in Eq.~\eqref{eq-Ph1D} is adjusted to match the results of the tight-binding simulation at large $L$. 
}
\label{fig: conductance oscillations}
\end{figure}

For illustration, in Fig.~\ref{fig: conductance oscillations} we show the downstream conductance as a function of the length of the QH-SC interface obtained from a full tight-binding simulation of the structure shown in Fig.~\ref{fig: qh_sc_junction}.
Here the same parameters were used as in Fig.~\ref{fig: tau vs theta non-ideal interface} with $\nu = 2.8$.
It is compared with the result of an effective 1D model where we set $L_b = \xi/10$ with $\xi = v_{F}^{SC}/\Delta$ the BCS coherence length, $v_b = v_{CAES}$ and $\mu_{b} = \mu_{SC}$. 
{We use a numerical minimization procedure to find the values of $\Delta_b$ and $\phi_b$ that give the scattering probabilities $\tau_1,\tau_2$ obtained from the tight-binding model. Fitting parameters for Fig.~\ref{fig: conductance oscillations} were $\Delta_{b1}=10.08\Delta$, $\phi_{b1}=3.382$ and $\Delta_{b2}=0.32\Delta$, $\phi_{b2}=3.002$. (Note that the choice is not unique.) In addition, we adjust the scattering phase $\phi_{12}$ appearing in Eq.~\eqref{eq-Ph1D} so that the effective model matches the simulation at large $L$.} A small mismatch between the values of $k_0$ can be attributed to lattice effects. Furthermore, deviations are visible at small lengths when the two corners cannot be treated independently, as assumed in Eq.~\eqref{eq-Ph1D}.

\section{Further considerations\label{sec: limits of 1D model}}

 In addition to the difficulty of determining parameters, effective 1D models have other obvious limitations.

\begin{figure}\centering
\includegraphics[width=0.7\linewidth]{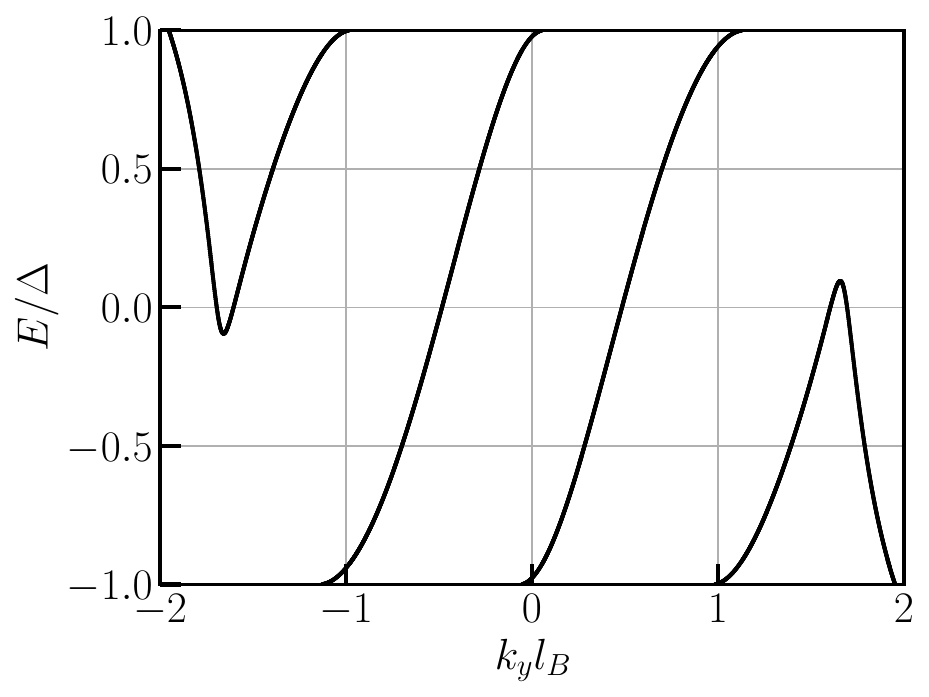}
\caption{Spectrum with additional nonchiral zero-energy edge states. Here we set $\nu=2.8$, $\mu_{QH} = \mu_{SC} = 20\Delta$ and $Z = 0$. These additional states appear for $\nu\lesssim 3$ and close to ideal interfaces.}
\label{fig: spectrum with track states}
\end{figure}

The effective 1D model only describes the topologically-protected chiral edge states. As can be seen in Fig.~\ref{fig: example of CAES spectrum}, in a full 2D calculation, additional subgap states may appear. While for the parameters chosen in Fig.~\ref{fig: example of CAES spectrum}, these states are close to the gap edge, they may cross the Fermi level in other parameter regimes. An example is shown in Fig.~\ref{fig: spectrum with track states}. We studied their parameter dependence and found zero-energy crossings only happen  for $\nu\lesssim 3$ and close to ideal interfaces. In experimentally relevant regimes, they are not expected to play a role as discussed in Appendix~\ref{app: track states}. 
Note that additional in-gap states may appear as well when the interface is smooth. This question has been addressed in Ref.~\cite{manesco2021}.

The downstream conductance at finite temperature is more likely affected by these nonchiral states. Furthermore, at finite $T$, the linear approximation for the dispersion of the CAES may not be sufficient. Namely, as long as $k_BT\ll\Delta$ and continuum contributions may be neglected, the downstream conductance $G_d(T)$ takes the form: 
\begin{align}
G_d (T)  \approx G_0 \int_{-\Delta}^\Delta  dE\frac{1- 2P_h(E)}{4k_BT\cosh^2\left(\frac{E}{2 k_B T}\right)},
\label{eq: finite temperature conductance}
\end{align}
where $P_h(E)$ is given by Eq.~\eqref{eq-Ph1D} by replacing $2k_0$ with $\delta k(E)=k_{qe}(E)-k_{qh}(E)$, where $k_{qe/qh}(E)=E/v\pm k_0$, and using the transmission probabilities $\tau_i$ at energy $E$. If $\delta k$ varies significantly with energy on the scale $k_BT$, this leads to an averaging of the oscillations of the downstream conductance. Numerically we find that the effect is small in experimentally relevant parameter regimes, see Appendix~\ref{app: finite temperature}.

\section{Conclusion\label{sec: conclusion}}

In this paper, we have studied the downstream conductance mediated by CAES in QH-SC junctions. In particular, we found that the geometry plays an important role. This limits the applicability of simple effective 1D models that are often used to describe such systems. We showed that the most general effective 1D model containing a complex pairing potential localized in the region where the QH-vacuum edge meets the QH-SC edge allows one to model an arbitrary electron-hole conversion probability---however, there is no clear prescription as to how parameters have to be chosen. We note that the geometry dependence may be exploited to device asymmetric junctions, where the overall electron-hole conversion probability is enhanced and the average downstream conductance can become negative. This may be a way to obtain clearer signatures of the Andreev conversion at the QH-SC interface. Our work concentrated on the clean case. It will be interesting to explore how these features are modified by disorder. Disorder as well as vortices modify the propagation phase along the interface and therefore change the interference pattern. However, this effect alone does not modify the minimal and maximal values, nor the average value of the downstream conductance determined by the electron-hole conversion at the corners. Thus, the geometrical effects are expected to be robust as long as the disorder does not introduce significant electron-hole scattering along the interface. On the other hand, vortices may lead to the loss of quasiparticles, thus decreasing the overall value of the downstream conductance. It is less clear how this affects the repartition between quasielectrons and quasiholes. Further studies are also needed to better characterize geometries with a narrow superconducting finger such that crossed Andreev reflections and cotunneling across the finger come into play.

\begin{figure}[!t]\centering
\subfloat[]{\includegraphics[width=0.33\linewidth]{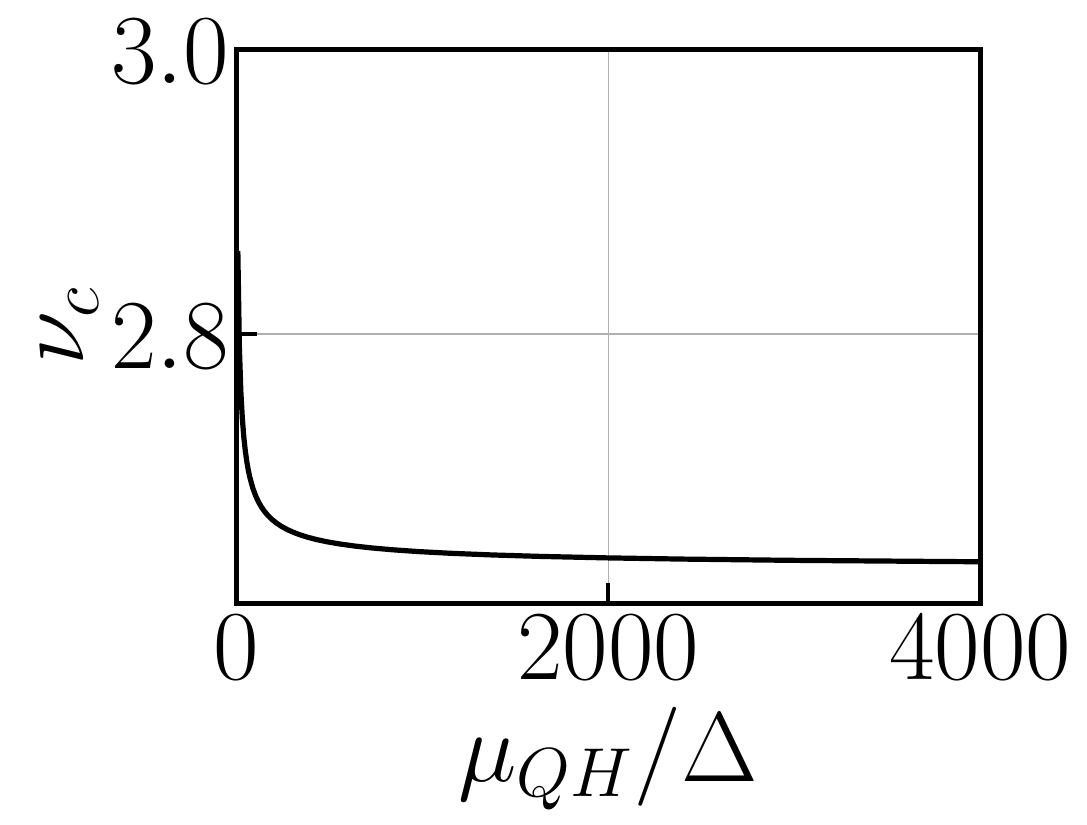}}
\subfloat[]{\includegraphics[width=0.33\linewidth]{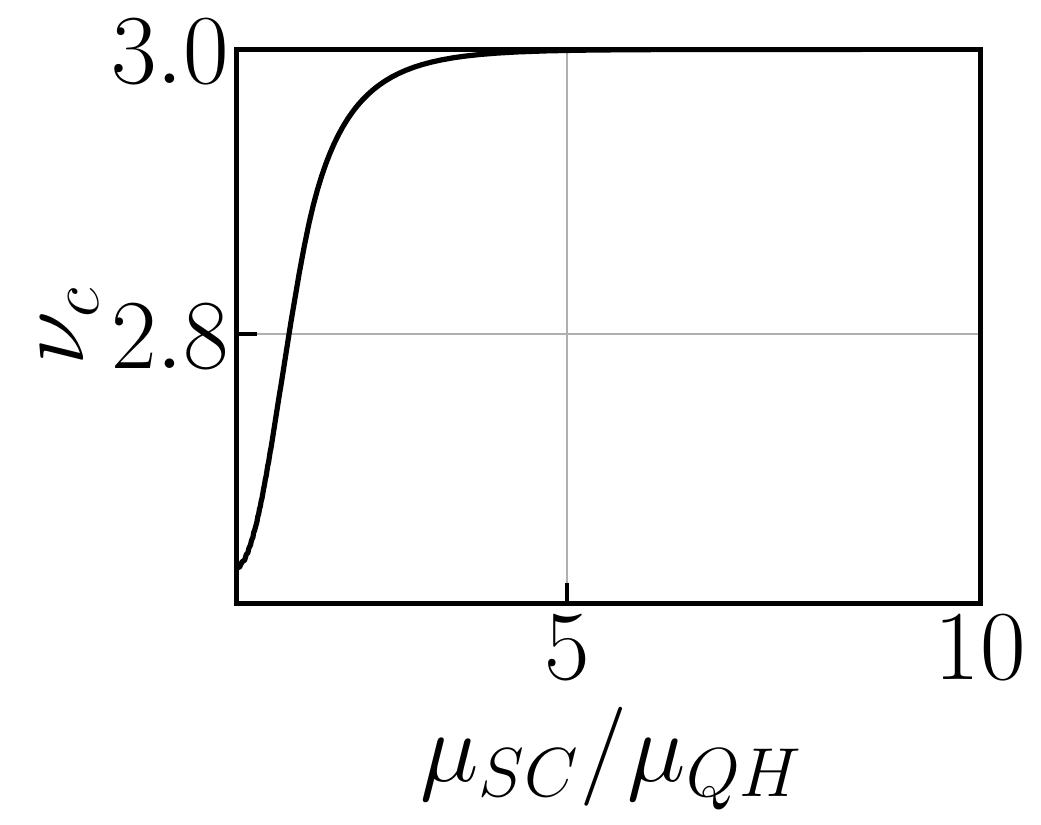}}
\subfloat[]{\includegraphics[width=0.33\linewidth]{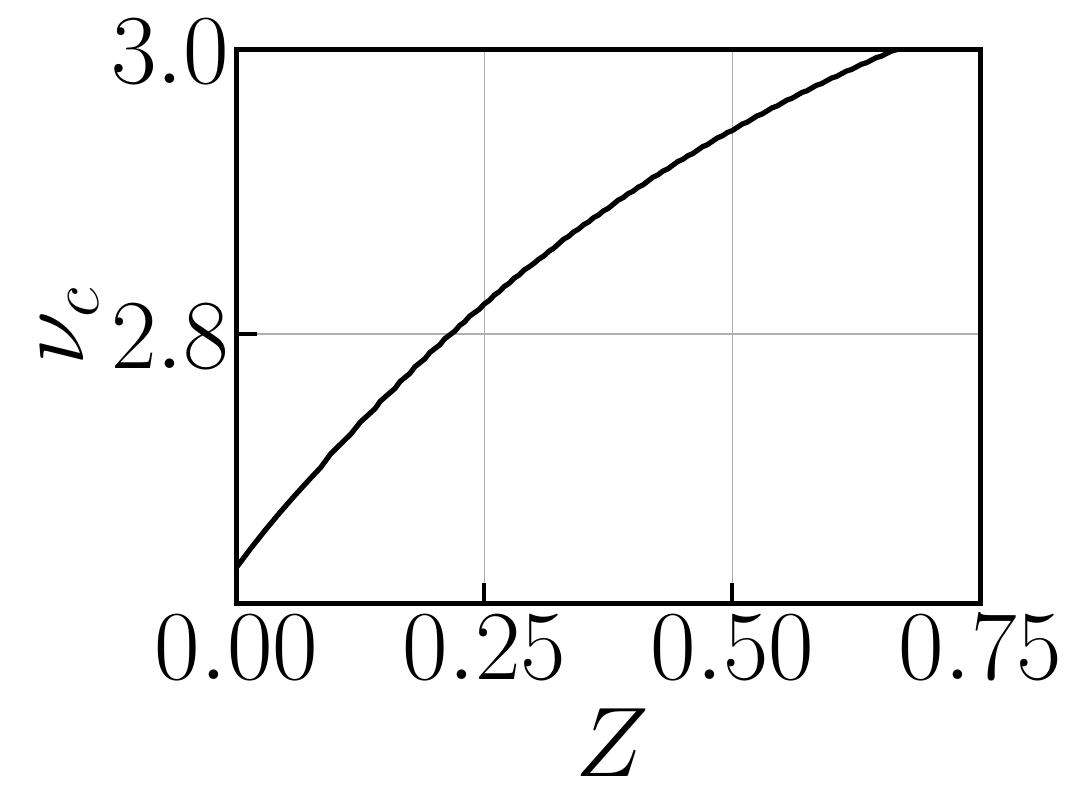}}
\caption{ {Plots of $\nu_{c}$, indicating the appearance of additional non-chiral edge states at the Fermi level, as a function of different parameters.
(a)  Dependence of $\nu_c$ on $\mu_{QH}/\Delta$ for an ideal interface, $\mu_{SC}=\mu_{QH}$ and $Z=0$.  In the limit $\mu_{QH}/\Delta\to\infty$, the critical value tends to {$\nu_c\approx2.63$}. (b) Dependence of $\nu_c$ on the mismatch $\mu_{SC}/\mu_{QH}$ at $\Delta=\mu_{QH}\times10^{-6}$ and $Z=0$. As $\nu_c$ reaches three, the additional nonchiral subgap states disappear at moderate values of the mismatch. (c) Dependence of $\nu_c$ on the barrier strength $Z$  at $\Delta=\mu_{QH}\times10^{-6}$ and $\mu_{SC}=\mu_{QH}$. As $\nu_c$ reaches three, the additional nonchiral subgap states disappear at moderate values of the barrier strength.}}
\label{fig: nu_crit limit non ideal interface}
\end{figure}

\begin{acknowledgments}
We thank X. Waintal for help with Kwant.
A.D. gratefully acknowledges interesting discussions with A. Bondarev. Furthermore, we acknowledge support from the French Agence Nationale de la Recherche (ANR) through Grants No.~ANR-17-PIRE-0001 and No.~ANR-21-CE30-0035.
\end{acknowledgments}

\appendix

\begin{figure}[!t]\centering
\subfloat[\label{fig: gap45}]{\includegraphics[width=0.4\linewidth]{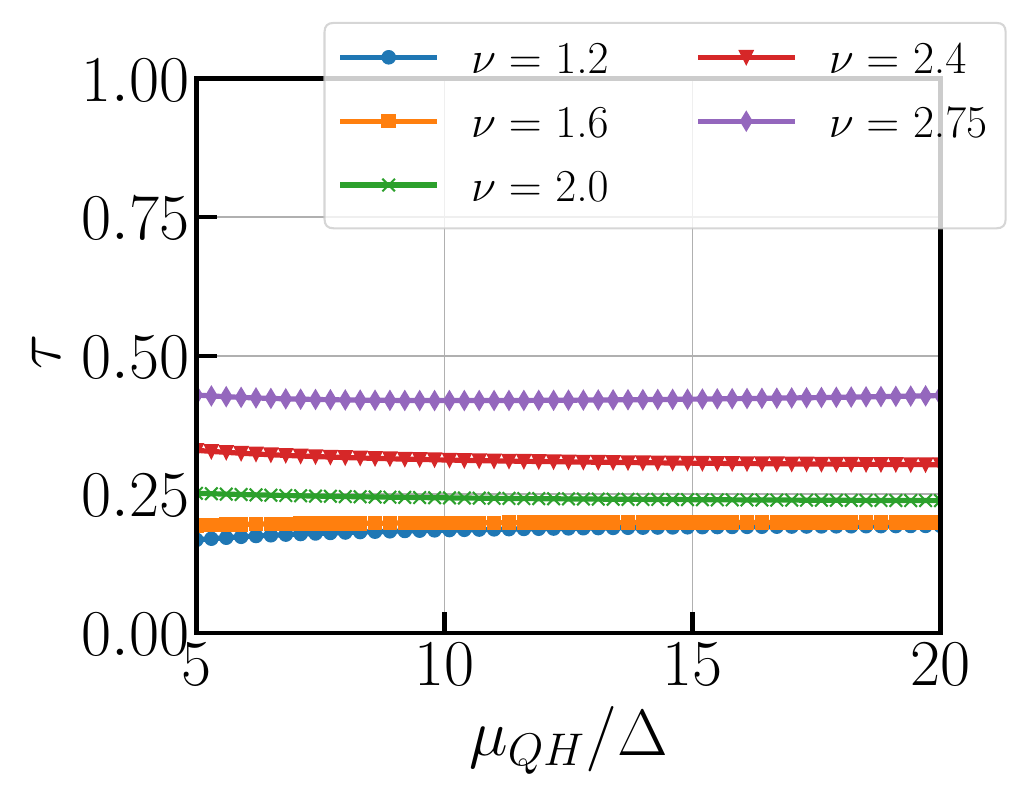}}
\subfloat[\label{fig: gap135}]{\includegraphics[width=0.4\linewidth]{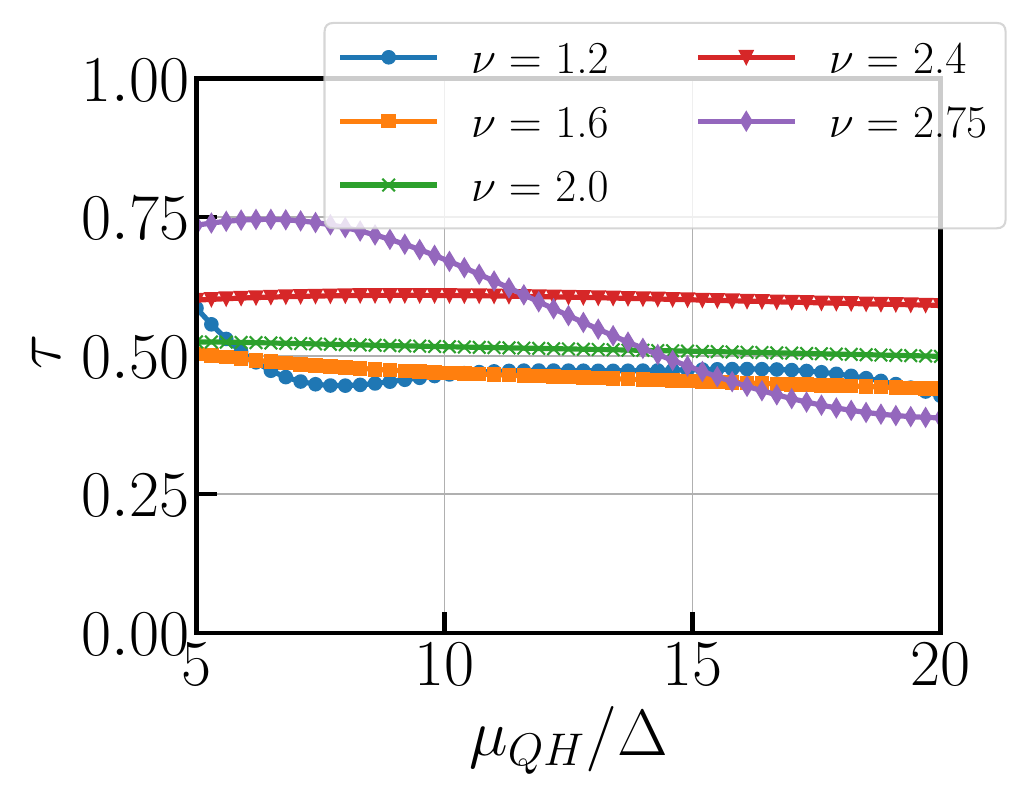}}
\caption{
Dependence of the electron-hole conversion probability $\tau$ on the superconducting gap $\Delta$. The parameters are $\mu_{QH} = \mu_{SC}$,  $Z=0$, and $\theta_{QH}=90^\circ$. (a) At $\theta_{SC}=45^\circ$, the electron-hole conversion probability very weakly depends on $\Delta$ in the regime $\Delta\ll\mu_{QH}$. (b)  At $\theta_{SC}=135^\circ$, a stronger dependence is seen. This can be attributed to the observation that, for angles $\theta_{SC}>90^\circ$, the superconductor-vacuum interface comes into play and may modify the decay as illustrated in Fig.~\ref{fig: sc-vac}.
}
\label{fig: gap}
\end{figure}

\begin{figure}[!b]\centering
\subfloat[\label{fig: decay135_1}]{\includegraphics[width=0.5\linewidth]{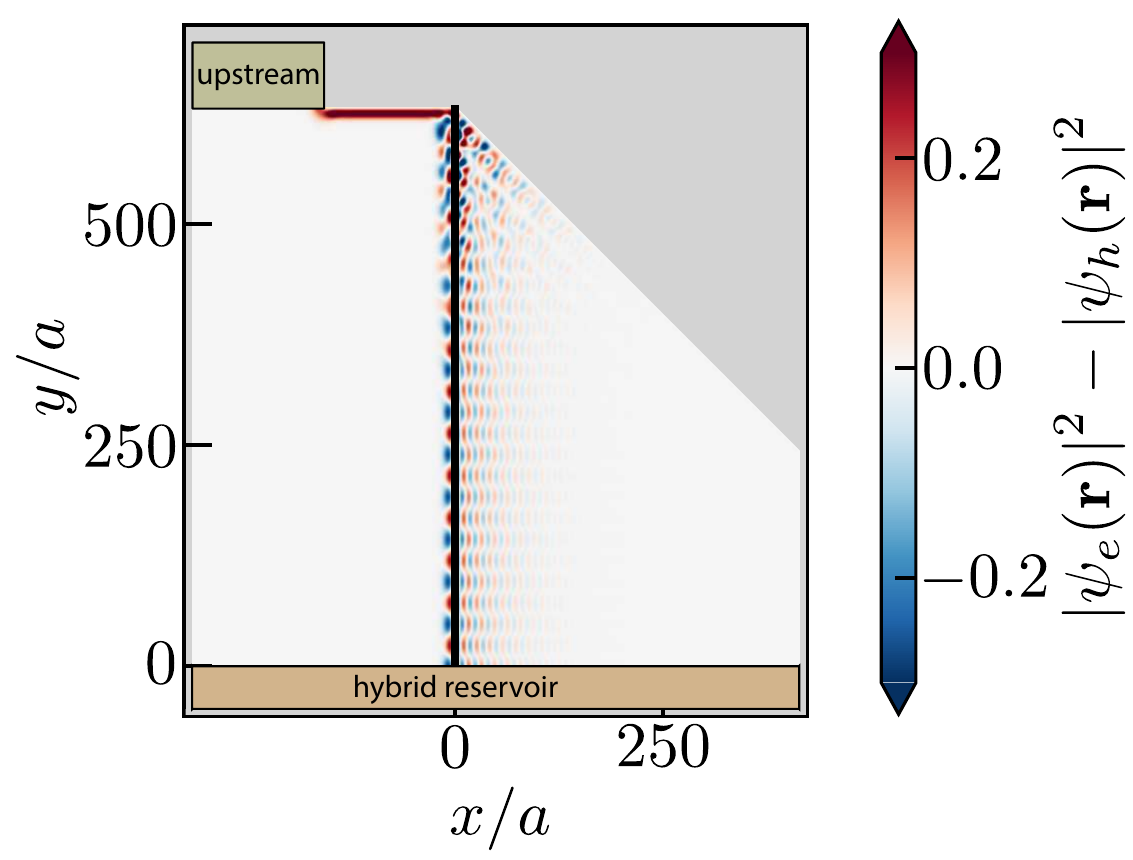}}
\subfloat[\label{fig: decay135_2}]{\includegraphics[width=0.5\linewidth]{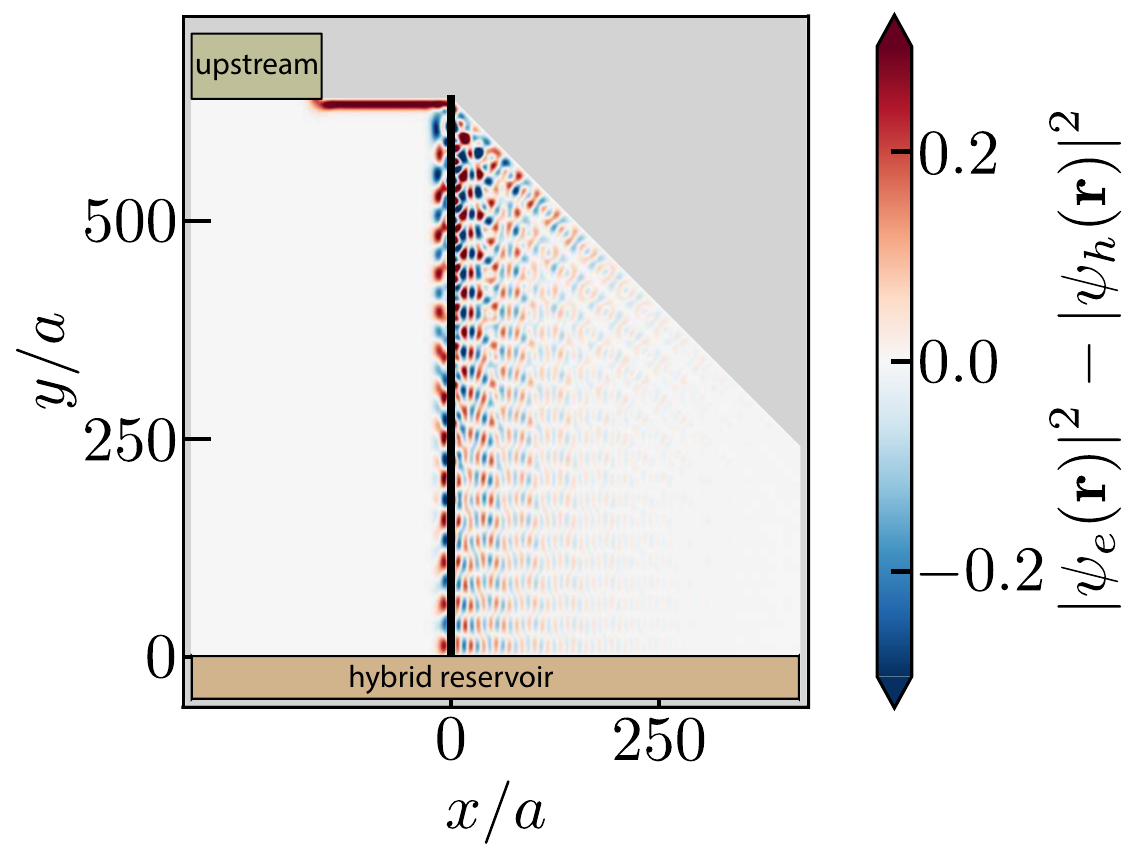}}
\caption{\label{fig: sc-vac}Probability density $|\psi_e(\mathbf{r})|^2 - |\psi_h(\mathbf{r})|^2$ of an incoming electron state for $\theta_{SC}=135^{\circ}$ and $\theta_{QH} = 90^{\circ}$.  Other parameters are $\nu=2.75$,  $\mu_{SC}=\mu_{QH}$, and $Z=0$.(a) $\mu_{QH}/\Delta=10$. (b) $\mu_{QH}/\Delta=20$.
The modified decay in the superconductor and the effect of the superconductor-vaccum interface can be clearly seen.}
\end{figure}

\begin{figure}[!t]\centering
\subfloat[\label{fig: energy spectrum non-ideal interface}]{\includegraphics[width=0.4\linewidth]{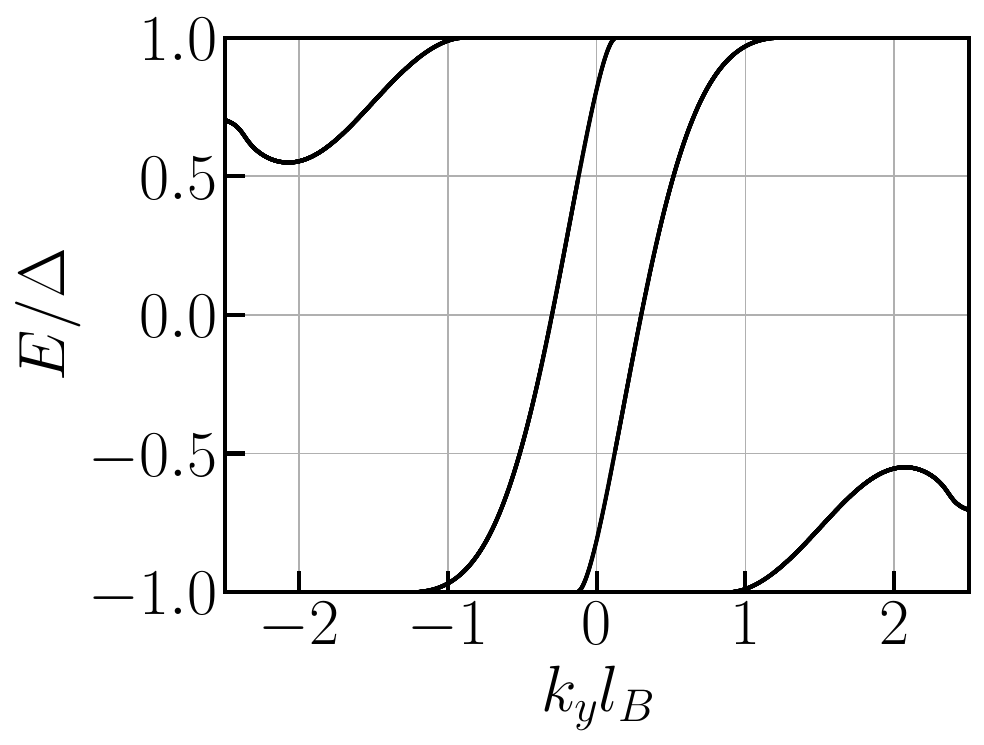}}\quad
\subfloat[\label{fig: momentum difference vs energy}]{\includegraphics[width=0.4\linewidth]{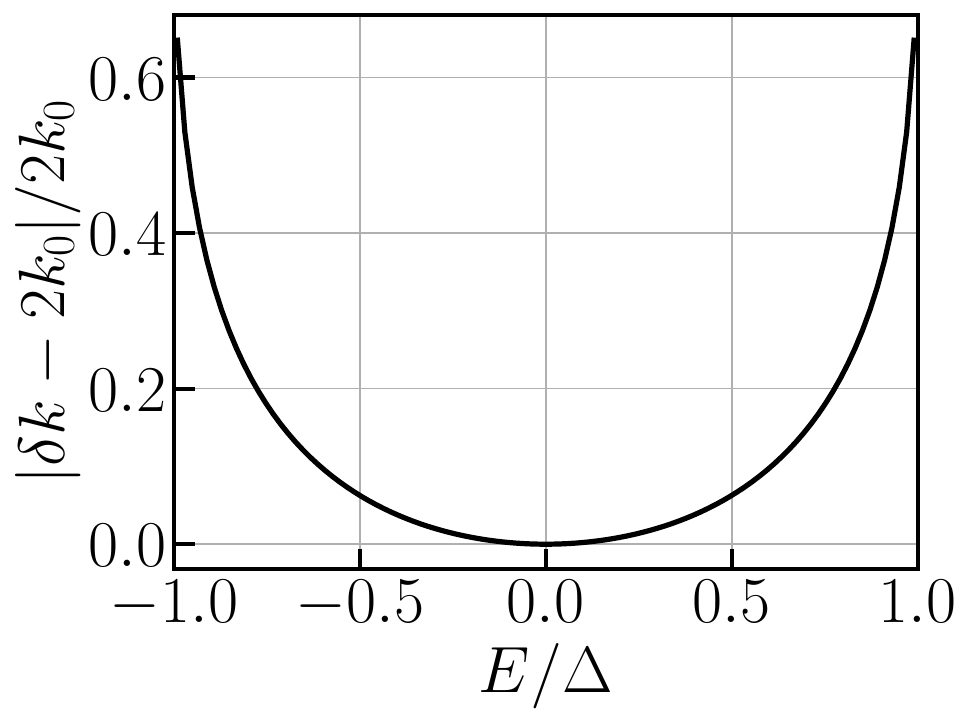}}\quad

\subfloat[\label{fig: tau vs energy theta_QH=0}]{\includegraphics[width=0.4\linewidth]{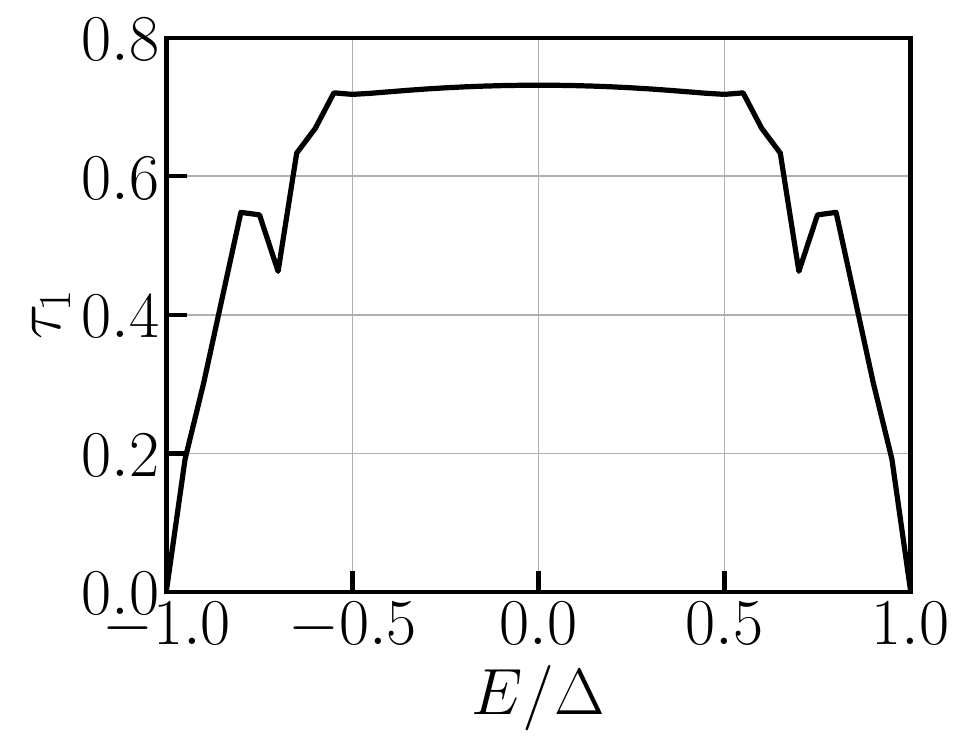}}
\subfloat[\label{fig: tau vs energy theta_QH=90}]{\includegraphics[width=0.4\linewidth]{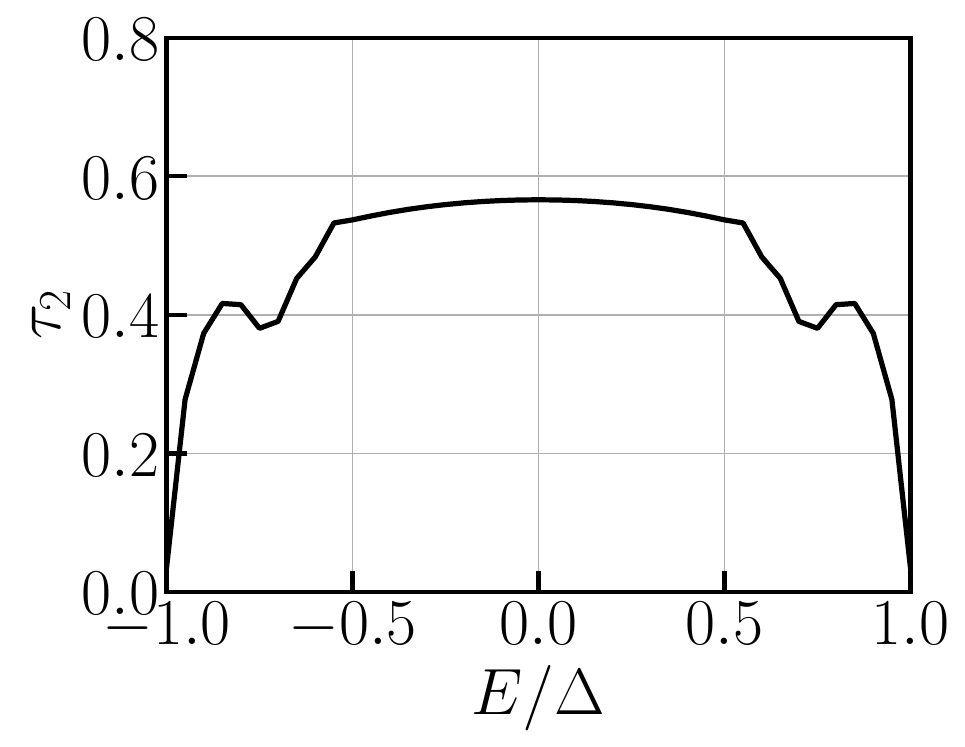}}
\caption{
{Energy dependence of various parameters necessary to determine  the downstream conductance at finite temperature. Parameters are the same as in Fig.~\ref{fig: conductance oscillations}. (a) Energy spectrum.  
(b) Variation of the relative momentum difference $|\delta k(E)-2k_0|/2k_0$ of the pair of CAES. 
(c) Conversion probability $\tau_1=\tau(\theta_{QH} = 0,\theta_{SC}=90^\circ)$ and (d) $\tau_2=\tau(\theta_{QH} = 90^\circ,\theta_{SC}=90^\circ)$. Variations in Figs.~(b)--(d) are seen to be small as long as $|E|\ll\Delta$.}
}
\label{fig: energy dependence of conductance parameters}
\end{figure}

\section{Additional non-chiral edge states\label{app: track states}}

As discussed in the main text, see Fig.~\ref{fig: spectrum with track states}, additional nonchiral edge states may cross the Fermi level in certain parameter regimes as $\nu$ approaches three. 
Using the continuous model of Sec.~\ref{ssec:2Dc}, we may determine the value $\nu_c$ above which such states are present as a function of system parameters. 
To do so we need to solve the secular equation, Eq.~\eqref{eq: secular equation}, at $E=0$ and determine the value $\nu_c$ at which a second solution with $k_y>k_0$ appears. The results are shown in Fig.~\ref{fig: nu_crit limit non ideal interface} as a function of $\mu_{QH}/\Delta$ for an ideal interface as well as a function of the interface barrier strength $Z$ and mismatch $\mu_{SC}/\mu_{QH}$ at $\Delta=\mu_{QH}\times10^{-6}$. 
We see that at $\Delta/\mu_{QH}\ll1$, additional zero-energy states appear for {$\nu>\nu_c\approx 2.63$ in the case of an ideal interface. An interface barrier, as well as potential mismatch, push that critical value up. It reaches three at $\mu_{SC}/\mu_{QH}\gtrapprox 3.73$ or $Z \gtrapprox 0.65$}. Beyond these values, one never finds additional zero-energy states, which is likely the case in experiments.


\section{Dependence of the electron-hole conversion probability on the superconducting gap\label{app: gap}}

In the main text, we show the dependence of the electron-hole conversion probability at the corners on various parameters. Here we complement our study with results on the dependence on the superconducting gap $\Delta$. In particular, we compare two different geometries, namely $\theta_{QH}=90^\circ$, $\theta_{SC}=45^\circ$ in Fig.~\ref{fig: gap45} and $\theta_{QH}=90^\circ$, $\theta_{SC}=135^\circ$ in Fig.~\ref{fig: gap135}. As shown in Appendix~\ref{app: track states}, nonchiral edge states may appear upon decreasing $\Delta$. Here we restrict ourselves to values of $\nu$ such that these states are absent in the range of values of $\Delta$ plotted. (In particular, we show results for $\nu=2.75$ rather than $\nu=2.8$ as in the main text.) For $\theta_{SC}=45^\circ$ (Fig.~\ref{fig: gap45}), the electron-hole conversion probability depends on $\Delta$ only very weakly. This is consistent with the analytic results of Sec.~\ref{ssec:2Dc}, which show that the properties of the edge states are almost independent of $\Delta$ in the considered parameter regime.
For $\theta_{SC}=135^\circ$ (Fig.~\ref{fig: gap135}), a stronger dependence is seen, in particular for $\nu$ close to one and three. For angles $\theta_{SC}>90^\circ$, the decay length of the edge state in the superconductor  plays a more important role. Namely as the decay may reach the superconductor-vacuum interface, a stronger dependence of $\tau$ on $\Delta$, which controls the decay length in the superconductor, is expected. The modified decay is illustrated in Fig.~\ref{fig: sc-vac}.

\begin{figure}[!b]\centering
\includegraphics[width=0.7\linewidth]{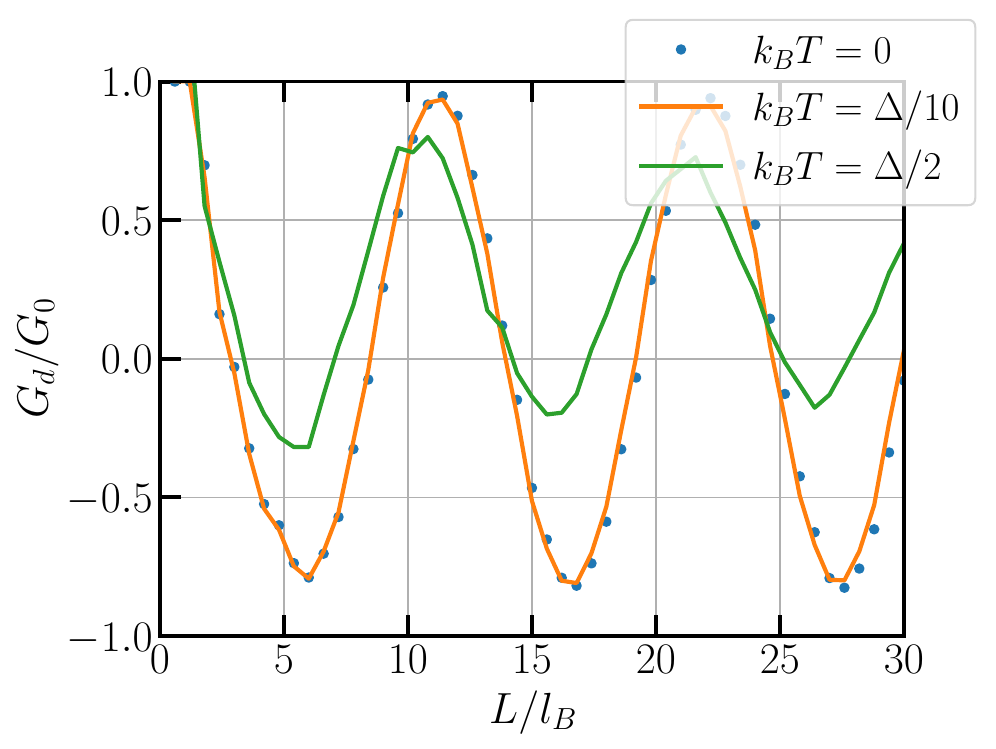}
\caption{Downstream conductance at different temperatures. The zero-temperature result is shown by blue dots. At $k_BT = \Delta/10$ (orange line), there is almost no change. By contrast, a clear reduction of the amplitude of the oscillations is observed at $k_BT=\Delta/2$ (green line). Parameters are the same as in Fig.~\ref{fig: conductance oscillations}.}
\label{fig: finite temperature conductance}
\end{figure}

\section{Downstream conductance at finite temperature\label{app: finite temperature}}

{As the downstream conductance at finite temperature involves integral over the hole conversion probabilities at different energies, it is important to know the dependence of the parameters determining the hole conversion probability on energy. In particular, if the momentum mismatch $\delta k$ or the phase $\phi_{12}$ strongly varies with energy,  the oscillations of the conductance should be averaged out upon increasing temperature.}

The momentum mismatch $\delta k(E)$ can be obtained from the continuum model. We find that, even beyond the regime where the edge state spectrum is linear, the variation of $\delta k$ remains small. 
We illustrate our findings in Fig.~\ref{fig: energy dependence of conductance parameters}. Here the same parameters as in Fig.~\ref{fig: conductance oscillations} were used. The spectrum is shown in Fig.~\ref{fig: energy spectrum non-ideal interface}. Additional non-chiral edge states are visible at energies $|E|\gtrsim \Delta/2$. The relative deviations of $\delta k(E)$ from $\delta k(0)=2k_0$ are shown in Fig.~\ref{fig: momentum difference vs energy}. For small enough energies, the deviations are small, implying a nearly constant period of the oscillations. Figures \ref{fig: tau vs energy theta_QH=0} and \ref{fig: tau vs energy theta_QH=90} show the energy dependence of the conversion probabilities $\tau_1$ and $\tau_2$. Again, the variation is weak up to the energy where additional subgap states appear.  Note that this is consistent with what one would obtain from our effective 1D model, where there is no energy dependence. {The scattering phase $\phi_{12}$ (not shown) remains approximatively constant in this regime as well.} These findings suggest that the zero-temperature results obtained for the downstream conductance are robust as long as $k_BT\ll\Delta$. This is confirmed by a full tight-binding simulation, shown in Fig.~\ref{fig: finite temperature conductance}. For $k_BT/\Delta=0.1$, the result is almost unaffected. By contrast, at the larger temperature $k_BT/\Delta=0.5$, a clear suppression of the amplitude of the oscillations is observed  {while the mean value increases as  energies close to $\Delta$ start to contribute, where variations of $\delta k$ become nonnegligible and $\tau_i\to0$}.

%
%

\begin{thebibliography}{99}

\bibitem{takagaki1998}
Y. Takagaki, 
Transport properties of semiconductor-superconductor junctions in quantizing magnetic fields,
\href{https://journals.aps.org/prb/abstract/10.1103/PhysRevB.57.4009}{Phys. Rev. B {\bf 57}, 4009 (1998)}.

\bibitem{asano2000}
Y. Asano and T. Kato, 
Andreev reflection and cyclotron motion of a quasiparticle in high magnetic fields,
\href{https://journals.jps.jp/doi/10.1143/JPSJ.69.1125}{J. Phys. Soc. Jpn. {\bf 69}, 1125 (2000)}.

\bibitem{chtchelkatchev2001}
N. M. Chtchelkatchev, 
Conductance of a semiconductor(2DEG)-superconductor junction in high magnetic field,
\href{https://link.springer.com/article/10.1134/1.1358428}{JETP Lett. {\bf 73}, 94 (2001)}.

\bibitem{chtchelkatchev2007}
N. M. Chtchelkatchev and I. S. Burmistrov,
Conductance oscillations with magnetic field of a two-dimensional electron gas–superconductor junction,
\href{https://journals.aps.org/prb/abstract/10.1103/PhysRevB.75.214510}{Phys. Rev. B {\bf 75}, 214510 (2007)}.

\bibitem{hoppe2000}
H. Hoppe, U. Zülicke, and G. Schön,
Andreev reflection in strong magnetic fields,
\href{https://journals.aps.org/prl/abstract/10.1103/PhysRevLett.84.1804}{Phys. Rev. Lett. {\bf 84}, 1804 (2000)}.

\bibitem{zulicke2001}
U. Zülicke, H. Hoppe, and G. Schön,
Andreev reflection at superconductor–semiconductor interfaces in high magnetic fields,
\href{https://www.sciencedirect.com/science/article/abs/pii/S0921452601003611}{Physica B {\bf 298}, 453 (2001)}.

\bibitem{giazotto2005}
F. Giazotto, M. Governale, U. Zülicke, and F. Beltram,
Andreev reflection and cyclotron motion at superconductor—normal-metal interfaces,
\href{https://journals.aps.org/prb/abstract/10.1103/PhysRevB.72.054518}{Phys. Rev. B {\bf 72}, 054518 (2005)}.

\bibitem{khaymovich2010}
I. M. Khaymovich, N. M. Chtchelkatchev, I. A. Shereshevskii, and A. S. Mel’nikov,
Andreev transport in two-dimensional normal-superconducting systems in strong magnetic fields,
\href{https://iopscience.iop.org/article/10.1209/0295-5075/91/17005}{Europhys. Lett. {\bf 91}, 17005 (2010)}.

\bibitem{ostaay2011}
J. A. M. van Ostaay, A. R. Akhmerov, and C. W. J. Beenakker,
Spin-triplet supercurrent carried by quantum Hall edge states through a josephson junction,
\href{https://journals.aps.org/prb/abstract/10.1103/PhysRevB.83.195441}{Phys. Rev. B {\bf 83}, 195441 (2011)}.

\bibitem{nayak2008}
C. Nayak, S. H. Simon, A. Stern, M. Freedman, and S. D. Sarma,
Non-abelian anyons and topological quantum computation,
\href{https://journals.aps.org/rmp/abstract/10.1103/RevModPhys.80.1083}{Rev. Mod. Phys. {\bf 80}, 1083 (2008)}.

\bibitem{mong2014}
R. S. K. Mong, D. J. Clarke, J. Alicea, N. H. Lindner, P. Fendley, C. Nayak, Y. Oreg, A. Stern, E. Berg, K. Shtengel, and M. P. A. Fisher,
Universal topological quantum computation from a superconductor-abelian quantum Hall heterostructure,
\href{https://journals.aps.org/prx/abstract/10.1103/PhysRevX.4.011036}{Phys. Rev. X {\bf 4}, 011036 (2014)}.

\bibitem{clarke2014}
D. J. Clarke, J. Alicea, and K. Shtengel,
Exotic circuit elements from zero-modes in hybrid superconductor–quantum-Hall systems,
\href{https://www.nature.com/articles/nphys3114}{Nat. Phys. {\bf 10}, 877 (2014)}.

\bibitem{lee2017}
G. H. Lee, K. F. Huang, D. K. Efetov, D. S. Wei, S. Hart, T. Taniguchi, K. Watanabe, A. Yacoby, and P. Kim,
Inducing superconducting correlation in quantum Hall edge states,
\href{https://www.nature.com/articles/nphys4084}{Nat. Phys. {\bf 13}, 693 (2017)}.

\bibitem{zhao2020}
L. Zhao, E. G. Arnault, A. Bondarev, A. Seredinski, T. F. Q. Larson, A. W. Draelos, H. Li, K. Watanabe, T. Taniguchi, F. Amet, H. U. Baranger, and G. Finkelstein,
Interference of chiral Andreev edge states,
\href{https://www.nature.com/articles/s41567-020-0898-5}{Nat. Phys. {\bf 16}, 862 (2020)}.

\bibitem{gul2022}
Ö. Gül, Y. Ronen, S. Y. Lee, H. Shapourian, J. Zauberman, Y. H. Lee, K. Watanabe, T. Taniguchi, A. Vishwanath, A. Yacoby, and P. Kim,
Andreev reflection in the fractional quantum Hall state,
\href{https://journals.aps.org/prx/abstract/10.1103/PhysRevX.12.021057}{Phys. Rev. X  {\bf 12}, 021057 (2022)}.

\bibitem{zhao2022}
L. Zhao, Z. Iftikhar, T. F. Q. Larson, E. G. Arnault, K. Watanabe, T. Taniguchi, F. Amet, and G. Finkelstein,
Loss and decoherence at the quantum Hall-superconductor interface,
\href{https://arxiv.org/abs/2210.04842}{arXiv:2210.04842 (2022)}.

\bibitem{hatefipour2022}
M. Hatefipour, J. J. Cuozzo, J. Kanter, W. M. Strickland, C. R. Allemang, T. M. Lu, E. Rossi, and J. Shabani,
Induced superconducting pairing in integer quantum Hall edge states,
\href{https://pubs.acs.org/doi/10.1021/acs.nanolett.2c01413}{Nano Lett.  {\bf 22}, 6173 (2022)}.

\bibitem{manesco2021}
A. L. R. Manesco, I. M. Flór, C. X. Liu, and A. R. Akhmerov,
Mechanisms of Andreev reflection in quantum Hall graphene,
\href{https://scipost.org/SciPostPhysCore.5.3.045}{SciPost Phys. Core {\bf 5}, 045 (2022)}.

\bibitem{kurilovich2022}
V. D. Kurilovich, Z. M. Raines, and L. I. Glazman,
Disorder in Andreev reflection of a quantum Hall edge,
\href{https://arxiv.org/abs/2201.00273}{arXiv:2201.00273 (2022)}.

\bibitem{schiller2022}
N. Schiller, B. A. Katzir, A. Stern, E. Berg, N. H. Lindner, and Y. Oreg,
Interplay of superconductivity and dissipation in quantum Hall edges,
\href{https://arxiv.org/abs/2202.10475}{arXiv:2202.10475 (2022)}.

\bibitem{blonder1982}
G. E. Blonder, M. Tinkham, and T. M. Klapwijk,
Transition from metallic to tunneling regimes in superconducting microconstrictions: Excess current, charge imbalance, and supercurrent conversion,
\href{https://journals.aps.org/prb/abstract/10.1103/PhysRevB.25.4515}{Phys. Rev. B {\bf 25}, 4515 (1982)}.

\bibitem{abramowitz1972}
M. Abramowitz and I. A. Stegun, 
\textit{Handbook of Mathematical Functions} (Dover Publications, New York, 1972)

\bibitem{kulik1970}
I. O. Kulik,
Macroscopic quantization and the proximity effect in S-N-S junctions, 
Zh. Eksp. Teor. Fiz. {\bf 57}, 1745 (1969)
[\href{http://www.jetp.ras.ru/cgi-bin/e/index/e/30/5/p944?a=list}{Sov. Phys. JETP {\bf 30}, 944 (1970)}].

\bibitem{groth2014}
C. W. Groth, M. Wimmer, A. R. Akhmerov, and X. Waintal,
Kwant: a software package for quantum transport,
\href{https://iopscience.iop.org/article/10.1088/1367-2630/16/6/063065}{New J. Phys. {\bf 16}, 063065 (2014)}.

\bibitem{datta1995}
S. Datta, 
\textit{Electronic transport in mesoscopic systems} (Cambridge University Press, Cambridge, 1995)

\bibitem{david2023}
A. David, 
Geometrical effects on the downstream conductance in quantum-Hall--superconductor hybrid systems (code),
\href{https://doi.org/10.5281/zenodo.7627946}{Zenodo (2023)}, \url{https://doi.org/10.5281/zenodo.7627946}.


\end{thebibliography}

\end{document}